\documentclass[11pt]{article}
\usepackage{amssymb}
\usepackage{graphics}
\usepackage{epsfig}
\usepackage{a4wide}
\usepackage{color}

\begin{document}

\newcommand{\ppww}{$pp \to W^+ W^-+X~$ }
\newcommand{\ppwwg}{$pp \to W^+ W^-g+X~$ }
\newcommand{\ppwwpol}{$pp \to W^+_{\lambda_3}W^-_{\lambda_4}+X~$}
\newcommand{\qqww}{$q\bar q \to W^+W^-~$}
\newcommand{\qqwwpol}{$q\bar q \to W^+_{\lambda_3}W^-_{\lambda_4}~$}
\newcommand{\qqwwg}{$q\bar q \to W^+W^-g~$}
\newcommand{\ggww}{$gg \to W^+W^-~$}
\newcommand{\ggwwpol}{$gg \to W^+_{\lambda_3}W^-_{\lambda_4}~$}
\newcommand{\ggwwg}{$gg \to W^+W^-g~$}
\newcommand{\qgwwq}{$q(\bar q)g \to W^+W^-q(\bar q)~$}
\newcommand{\ppuuww}{$pp (u\bar u) \to W^+W^-+X~$}
\newcommand{\ppqqww}{$pp (q\bar q) \to W^+W^-+X~$}
\newcommand{\ppggww}{$pp (gg) \to W^+W^-+X~$}
\newcommand{\ppuuwwg}{$pp (u\bar u) \to W^+W^-g+X~$}

\title{ $WWZ/\gamma$ production in the large extra dimensions model at the LHC and ILC}
\author{ \textcolor{red}{}
Li Xiao-Zhou, Duan Peng-Fei, Ma Wen-Gan, Zhang Ren-You, and Guo Lei\\
{\small Department of Modern Physics, University of Science and Technology}\\
{\small of China (USTC), Hefei, Anhui 230026, P.R.China}  }

\date{}
\maketitle \vskip 15mm
\begin{abstract}
We investigate the effect induced by the Kaluza-Klein (KK) graviton
in the $W^+W^-\gamma/Z$ production in the framework of the large
extra dimensions (LED) model at both the CERN Large Hadron Collider
(LHC) and the International Linear Collider (ILC). The integrated
cross sections and various kinematic distributions in the LED model
are presented and compared with those in the standard model. The
results show that the contributions from KK-graviton exchange
remarkably affect the observables of the triple gauge boson
($W^+W^-\gamma/Z$) production processes at both the ILC and LHC,
particularly either in the high transverse momentum region or in the
central rapidity region. We also find that the relative LED
discrepancy for the $W^+W^-\gamma/Z$ production at the LHC is
generally larger than that at the ILC due to the additional LED
contribution via $gg$ fusion subprocess and the KK-graviton
exchanging resonant effect induced by the continuous large colliding
energy in $pp$ collision. We conclude that the $W^{+}W^{-}\gamma$
and $W^{+}W^{-}Z$ productions at the LHC could have the distinct
advantage over at the ILC from the aspect of effectively exploring
the LED signal in measuring $W^+W^-\gamma/Z$ production.
\end{abstract}


\vfill \eject

\baselineskip=0.32in

\renewcommand{\theequation}{\arabic{section}.\arabic{equation}}
\renewcommand{\thesection}{\Roman{section}.}
\newcommand{\nb}{\nonumber}

\newcommand{\Dir}{\kern -6.4pt\Big{/}}
\newcommand{\Dirin}{\kern -10.4pt\Big{/}\kern 4.4pt}
\newcommand{\DDir}{\kern -7.6pt\Big{/}}
\newcommand{\DGir}{\kern -6.0pt\Big{/}}

\makeatletter      
\@addtoreset{equation}{section}
\makeatother       

\vskip 5mm
\section{Introduction}
\par
The CERN Large Hadron Collider (LHC) and the upcoming International
Linear Collider (ILC) are expected to perform precision tests of the
standard model (SM) and explore the new physics at the TeV
scale \cite{1}. The large extra dimensions (LED) model is one of the
scenarios beyond the SM which are proposed to solve the hierarchy
problem \cite{2}. The LED model has only one fundamental scale, $M_S
\sim $~TeV, and it may induce predictable collider phenomena at both the
LHC and the ILC. Up to now, many works on both the virtual
Kaluza-Klein (KK) graviton exchange and the real KK-graviton
production have been presented; for example, the $e^+ e^- \to V V$
and $p p \to V V,~V G,~G+jet$ processes were studied in the LED
model in Refs.\cite{3}-\cite{6}.

\par
In fact, the triple gauge boson (TGB) production processes are
sensitive to the quartic gauge couplings (QGCs) and thus
related to the electroweak symmetry breaking mechanism
\cite{9}. Any deviation from the SM prediction hints at the existence
of new physics, such as the Higgsless or extra dimension signals
\cite{9,10}. In discriminating physics beyond the SM, we should
investigate the potential contributions from the extension
models. Compared with the thoroughly studied diboson production
processes in extra dimension models, the TGB productions have been
fully studied in the SM \cite{7} but gained less attention in the
LED model. Not long ago, the neutral TGB production processes at the
LHC, $pp \to \gamma\gamma\gamma$, $pp \to \gamma\gamma Z$, $pp \to
\gamma ZZ$ and $pp \to ZZZ$, were studied in the framework of
the LED model in Ref.\cite{8}.

\par
In this paper, we investigate the possible contributions of the
virtual KK-graviton exchange to the $W^+W^-\gamma$ and $W^+W^-Z$
productions at both the LHC and ILC. The motivation for this
work is shown in two fields: Firstly, the $W^+W^-\gamma$ and $W^+W^-Z$
processes are directly related to the SM QGCs, namely
$W^+W^-\gamma\gamma$, $W^+W^-\gamma Z$ and $W^+W^-ZZ$, which
are different from the absence of the neutral QGCs in the SM at the
tree level \cite{8}. Secondly, although the experimental precision is
limited by our understanding of strong QCD background, the LHC can
provide more precision measurements of the QGCs than the existing
data from LEP II and Tevatron searches due to its very high energy
and luminosity \cite{11,13}. Furthermore, in the future, the QGCs
can be further probed with higher precision at the ILC
due to its cleaner environment \cite{14}. In this sense, the LHC and
the upcoming ILC will provide complementary studies on the TGB
production channels. The paper is organized as follows: In section
II, we present the related theory of the LED model used in our
calculations. In section III, the calculation strategies are
presented. The numerical results and analyses for the $W^+W^-\gamma$
and $W^+W^-Z$ production processes at both colliders are
provided in Section IV, and a short summary is given in the last
section.

\vskip 5mm
\section{ Related theory }
\label{related theories}
\par
In the LED model \cite{2}, the spacetime is $D=4+n$ with $n$ being
the number of extra dimensions. The only fundamental scale
$M_S$ unifying the gravity and the gauge interactions is at the
TeV. To explore the phenomenological effects, one can extract the
low-energy effective theory by the KK reduction in the brane-bulk
picture \cite{2,15}. In this scenario, the SM particles are confined
on a $(3+1)$-dimensional brane world volume while the gravity can
propagate in the $D$-dimensional bulk. After the assumed
torus compactification of the extra dimensions $n$, the usual Planck
scale $M_P$ in the $(3+1)$ spacetime is related to the fundamental
scale $M_S$ as $M_P^2 \sim R^n M_S^{n + 2}$ \cite{16}, where $R$ is
the radius of the $n$ torus.

\par
In the following calculations, we adopt the de Donder gauge for the
KK-graviton part, while the Feynman gauge ($\xi=1$) is used for the
SM part. We assume all the momenta flow to the vertices, except that
the fermionic momenta are set to be along with the fermion flow
directions. Then we list the Feynman rules for the relevant
vertices and the propagator of spin-2 KK-graviton in the LED model
below \cite{15}, where $G_{\rm KK}^{\mu \nu}$, $\psi$, $W^{\pm
\mu}$, $Z^{\mu}$ and $A^{\mu}$ represent the fields of the graviton,
fermion, $W$ boson, $Z$ boson and photon, respectively.
\begin{itemize}
\item
$G_{\rm KK}^{\mu
\nu}(k_3)-\bar{\psi}(k_1)-\psi(k_2)~\textrm{vertex}: $
\begin{eqnarray}
-i \frac{\kappa}{8} \left[\gamma^{\mu} (k_1 + k_2)^{\nu} +
\gamma^{\nu} (k_1 + k_2)^{\mu} - 2 \eta^{\mu \nu} (\rlap/{k}_1 +
\rlap/{k}_2 - 2 m_{\psi}) \right]
\end{eqnarray}
\item
$G_{\rm KK}^{\mu
\nu}(k_4)-\bar{\psi}(k_1)-\psi(k_2)-A^{\rho}(k_3)~\textrm{vertex}:
$
\begin{eqnarray}
i e Q_{f} \frac{\kappa}{4} \left( \gamma^{\mu} \eta^{\nu
\rho} + \gamma^{\nu} \eta^{\mu \rho} - 2 \gamma^{\rho}\eta^{\mu \nu}
\right)
\end{eqnarray}
\item
$G_{\rm KK}^{\mu
\nu}(k_4)-\bar{\psi}(k_1)-\psi(k_2)-Z^{\rho}(k_3)~\textrm{vertex}:
$
\begin{eqnarray}
-i e \frac{\kappa}{4} \left[(\gamma^{\mu} \eta^{\nu
\rho} + \gamma^{\nu} \eta^{\mu \rho} - 2 \gamma^{\rho}\eta^{\mu \nu})
(\upsilon_f -a_f \gamma_5) \right]
\end{eqnarray}
\item
$G_{\rm KK}^{\mu \nu}(k_3)-W^{+ \rho}(k_1)-W^{-
\sigma}(k_2)~\textrm{vertex}: $
\begin{eqnarray}
-i \kappa \left[B^{\mu \nu \rho \sigma} m_W^2 +
(C^{\mu \nu \rho \sigma \tau \beta} - C^{\mu \nu \rho \beta \sigma
\tau}) k_{1\tau} k_{2\beta} + \frac{1}{\xi}E^{\mu \nu \rho
\sigma}(k_1,k_2)\right]
\end{eqnarray}
\item
$G_{\rm KK}^{\mu \nu}(k_4)-W^{+ \rho}(k_1)-W^{-\sigma}(k_2)
-A^{\lambda}(k_3)~\textrm{vertex}: $
\begin{eqnarray}
-i e \kappa \left[(k_1-k_3)_{\tau}C^{\mu
\nu \tau \sigma \rho \lambda}+ (k_2-k_1)_{\tau}C^{\mu \nu \sigma
\rho \tau \lambda} + (k_3-k_2)_{\tau}C^{\mu \nu \lambda \sigma \tau
\rho}\right]
\end{eqnarray}
\item
$G_{\rm KK}^{\mu \nu}(k_4)-W^{+ \rho}(k_1)-W^{-\sigma}(k_2)
-Z^{\lambda}(k_3)~\textrm{vertex}: $
\begin{eqnarray}
i e \frac{s_w}{c_w} \kappa \left[(k_1-k_3)_{\tau}C^{\mu
\nu \tau \sigma \rho \lambda}+ (k_2-k_1)_{\tau}C^{\mu \nu \sigma
\rho \tau \lambda} + (k_3-k_2)_{\tau}C^{\mu \nu \lambda \sigma \tau
\rho}\right]
\end{eqnarray}
\end{itemize}
where $e=\sqrt{4\pi\alpha}$, $\alpha$ is the fine-structure
constant, $Q_f$ is the electric charge of fermion, $s_w~(c_w)$ are
sine (cosine) of the Weinberg angle, the vector and axial vector
couplings of the $Z$-boson, i.e., $\upsilon_f$ and $a_f$, are the
same as in the SM, and $\kappa=\sqrt{16\pi G_N}$ is related to the
reduced Planck mass as $\overline{M}_P=\sqrt{2}\kappa^{-1}$, where 
$G_N$ is the Newton constant. The tensor coefficients 
$B^{\mu \nu \alpha \beta}$, $C^{\rho \sigma \mu
\mu \alpha \beta}$ and $E^{\mu \nu \rho \sigma}(k_{1},k_{2})$ are
expressed as \cite{5-1}
\begin{eqnarray}
B^{\mu \nu \alpha \beta} & = & \frac{1}{2}
      (\eta^{\mu \nu}\eta^{\alpha \beta}
      -\eta^{\mu \alpha}\eta^{\nu \beta}
      -\eta^{\mu \beta}\eta^{\nu \alpha}),
       \nb \\
C^{\rho \sigma \mu \nu \alpha \beta} & = & \frac{1}{2}
      [\eta^{\rho \sigma}\eta^{\mu \nu}\eta^{\alpha \beta}
     -(\eta^{\rho \mu}\eta^{\sigma \nu}\eta^{\alpha \beta}
      +\eta^{\rho \nu}\eta^{\sigma \mu}\eta^{\alpha \beta}
      +\eta^{\rho \alpha}\eta^{\sigma \beta}\eta^{\mu \nu}
      +\eta^{\rho \beta}\eta^{\sigma \alpha}\eta^{\mu \nu})],
      \nb \\
E^{\mu \nu \rho \sigma}(k_{1},k_{2}) & = &
      \eta^{\mu \nu}(k_1^{\rho} k_1^{\sigma} + k_2^{\rho} k_2^{\sigma}
      + k_1^{\rho} k_2^{\sigma}) - \left [\eta^{\nu \sigma} k_1^{\mu} k_1^{\rho}
      + \eta^{\nu \rho} k_2^{\mu} k_2^{\sigma} + (\mu \leftrightarrow \nu)\right ]. \nb
\end{eqnarray}
After summation over KK states the spin-2 KK-graviton propagator can
be expressed as \cite{5-1}
\begin{eqnarray}
\tilde{G}_{\rm KK}^{\mu \nu \alpha \beta}=\frac{1}{2} D(s)
\left[\eta^{\mu \alpha} \eta^{\nu \beta} + \eta^{\mu \beta}
\eta^{\nu \alpha} - \frac{2}{n+2}\eta^{\mu \nu} \eta^{\alpha \beta}
\right],
\end{eqnarray}
where
\begin{equation} \label{Res}
D(s) = {{16\pi}\over{\kappa}^2} {s^{n/2-1}\over{M_S}^{n+2}}
\biggl[\pi + 2i I(\Lambda/\sqrt{s})\biggr],
\end{equation}
and
\begin{equation}
I(\Lambda/\sqrt{s}) = P \int_0^{\Lambda/\sqrt{s}}dy\ {y^{n-1}\over
1-y^2}.
\end{equation}
The integral $I(\Lambda/\sqrt{s})$ contains an ultraviolet cutoff
$\Lambda$ on the KK modes \cite{15,16}. It should be understood that
the point $y=1$ has been removed from the integration path, and we
set the ultraviolet cutoff $\Lambda$ to be the fundamental scale
$M_S$ routinely. The real part proportional to $\pi$ in
Eq.(\ref{Res}) is from the narrow resonant production of a single KK
mode with $m^2_{\vec{n}} = s$ and the imaginary part
$I(M_S/\sqrt{s})$ is from the summation over the many nonresonant
states.

\vskip 5mm
\section{ Calculations }
\label{calculations}
\par
\begin{figure*}
\begin{center}
\includegraphics[scale=0.8]{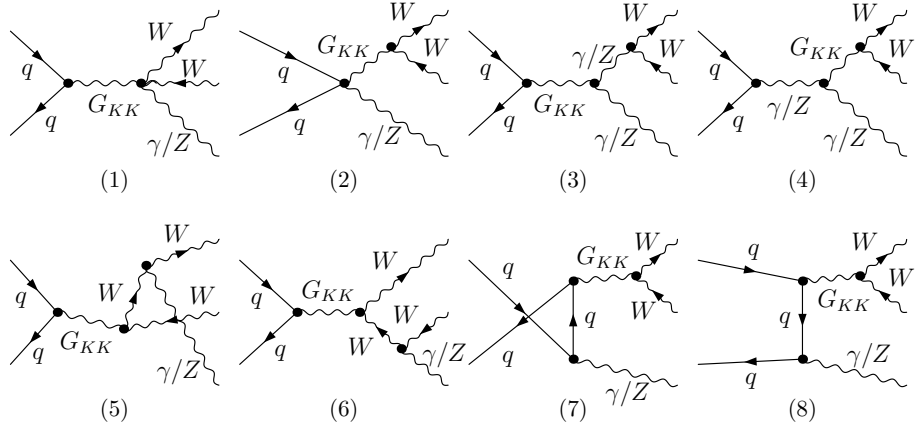}
\caption{\label{fig1} The LO Feynman diagrams for the partonic
process $q\bar q \to W^+W^-\gamma/Z$ with KK graviton exchange in
the LED model. The SM-like diagrams are not shown.}
\end{center}
\end{figure*}
\begin{figure*}
\begin{center}
\includegraphics[scale=0.8]{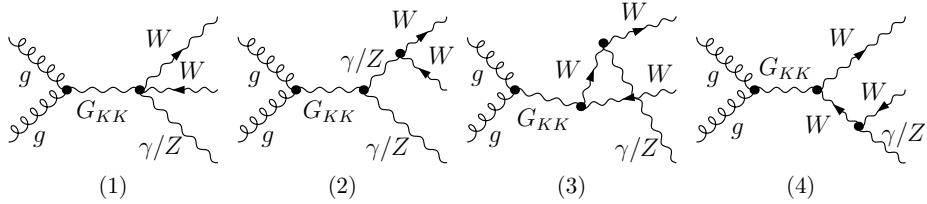}
\caption{\label{fig2} The LO Feynman diagrams for the gluon-gluon
fusion subprocess $g g \to W^+W^-\gamma/Z$ with KK-graviton exchange
in the LED model. }
\end{center}
\end{figure*}
\begin{figure*}
\begin{center}
\includegraphics[scale=0.8]{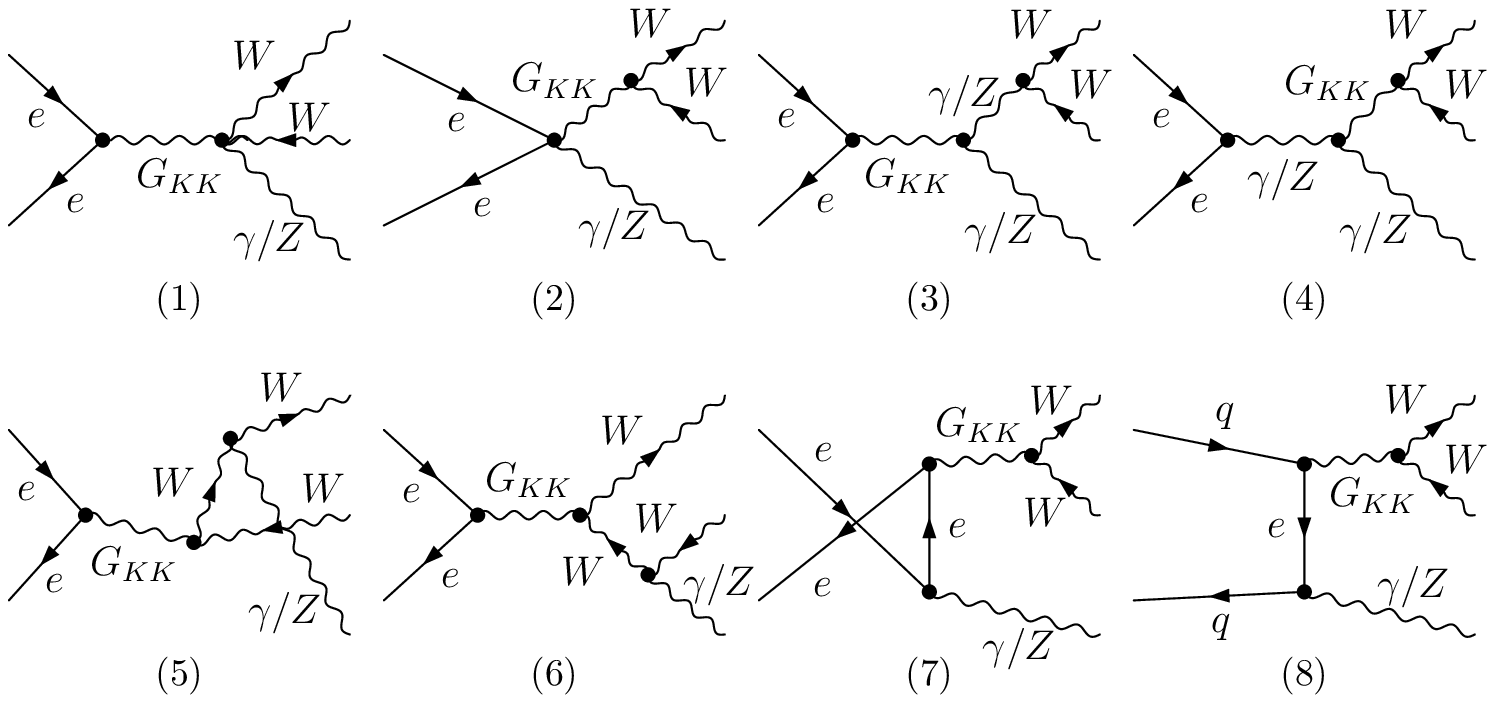}
\caption{\label{fig3} The LO Feynman diagrams for the $e^+e^- \to W^+W^-\gamma/Z$
process with KK-graviton exchange in the LED model. The SM-like diagrams are not shown.}
\end{center}
\end{figure*}

The $W^+W^-\gamma$ and $W^+W^-Z$ productions at
the LHC arise form the quark-antiquark annihilation
and the gluon-gluon fusion subprocesses at the parton level:
\begin{eqnarray}\label{channel-1}
~~~q(p_1)+\bar q(p_2) \to W^{+}(p_3)+W^{-}(p_4)+V(p_5),~~( q = u, d, s, c, b ) ,\\
\label{channel-2} g(p_1)+g(p_2) \to W^{+}(p_3)+W^{-}(p_4)+V(p_5).~~~~~~~~~~
\end{eqnarray}
The $e^+ e^- \to W^+W^-\gamma, W^+W^-Z$ processes at the ILC can be denoted as
\begin{eqnarray}\label{channel-3}
e^+(p_1)+e^-(p_2) \to  W^{+}(p_3)+W^{-}(p_4)+V(p_5).
\end{eqnarray}
In reactions (\ref{channel-1}), (\ref{channel-2}) and
(\ref{channel-3}), $V=\gamma,Z$, and $p_{i}$ $(i=1,2,3,4,5)$
represent the four-momenta of initial and final particles. The
leading order (LO) Feynman diagrams with KK-graviton exchange for
(\ref{channel-1}) and (\ref{channel-2}) channels are shown in
Figs.\ref{fig1} and \ref{fig2}, respectively, while the LO
additional Feynman diagrams in the LED model for the process
(\ref{channel-3}) are depicted in Fig.\ref{fig3}.

\par
From these Feynman diagrams, one can find that the KK-graviton
couples not only to the fermion pair, vector boson pair, and
fermion-antifermion-vector boson ($f\bar f V G_{KK}$), but also to the
TGB including charged gauge boson, which is absent in the neutral
TGB production processes as shown in Ref.\cite{8}. Therefore, it is
natural to expect that KK-graviton in the LED model may induce
considerable effects at the TeV scale on the TGB production
processes concerning charged gauge bosons at the LHC and the future
ILC.

\par
We express the Feynman amplitudes for the partonic processes $q
\bar{q} \to W^+W^-\gamma/Z$ and $gg \to W^+W^-\gamma/Z$ as
\begin{equation}
{\cal M}_{q\bar{q}}^{\gamma/Z} = {\cal M}_{q\bar{q}}^{\gamma/Z,SM} +
{\cal M}_{q\bar{q}}^{\gamma/Z,LED},~~~~{\cal M}_{gg}^{\gamma/Z} =
{\cal M}_{gg}^{\gamma/Z,LED},
\end{equation}
where ${\cal M}_{q\bar{q}}^{\gamma/Z,SM}$ $(q=u,d,c,s,b)$ is the
amplitude contributed by the SM-like diagrams, while ${\cal
M}_{q\bar{q}}^{\gamma/Z,LED}$ and ${\cal M}_{gg}^{\gamma/Z,LED}$ are
the amplitudes with KK-graviton exchange. The Feynman amplitude
for the $e^+ e^- \to W^+W^-\gamma/Z$ process can be expressed as
\begin{equation}
{\cal M}_{ee}^{\gamma/Z} = {\cal M}_{ee}^{\gamma/Z,SM} + {\cal
M}_{ee}^{\gamma/Z,LED}.
\end{equation}
where ${\cal M}_{ee}^{\gamma/Z,SM}$ stands for the amplitude
mediated by the SM-like particles, while ${\cal
M}_{ee}^{\gamma/Z,LED}$ is mediated by the KK-graviton.

\par
The total cross sections for the partonic process $q \bar{q}(gg) \to
W^+W^-\gamma/Z$ can be expressed as
\begin{eqnarray}\label{int-ppvvv}
\hat{\sigma}_{ij}^{\gamma/Z} =
\frac{1}{4|\vec{p}|\sqrt{\hat{s}}}\int {\rm d}\Gamma_3
\sum_{spin}{}^\prime\sum_{color}{}^\prime|{\cal M}_{ij}^{\gamma/Z}|^2,
~~~~(ij=u\bar{u},d\bar{d},c\bar{c},s\bar{s},b\bar{b},gg),
\end{eqnarray}
where $\vec{p}$ is the three-momentum of one initial parton in the
center-of-mass system (c.m.s), the summation is taken over the spins
or colors of initial and final particles, the prime on the sum
recalls averaging over initial spins or colors, and ${\rm d}
\Gamma_3$ is the three-body phase space element expressed as
\begin{eqnarray}
{\rm d} \Gamma_3 = (2\pi)^4  \delta^{(4)} \left(
p_1+p_2-\sum_{i=3}^5 p_i \right) \prod_{i=3}^5 \frac{d^3
\vec{p}_i}{(2\pi)^3 2E_i}.
\end{eqnarray}

\par
By convoluting $\hat{\sigma}^{\gamma/Z}_{i j}$ with the parton
distribution functions (PDFs) of the colliding protons, the total
cross section for the $pp \to W^+W^-\gamma/Z$ parent process
can be written as
\begin{eqnarray}
\sigma_{pp}^{\gamma/Z} = \sum^{c\bar c,b\bar b,gg}_{ij=u\bar
u,d\bar d,s\bar s}\frac{1}{1+\delta_{ij}} \int dx_A dx_B \left[
G_{i/A}(x_A,\mu_f)
G_{j/B}(x_B,\mu_f)\hat{\sigma}^{\gamma/Z}_{ij}(\sqrt{\hat{s}}=x_Ax_B\sqrt{s})
+(A \leftrightarrow B) \right], \nb \\
\end{eqnarray}
where $G_{i/P}$ $(i = q, \bar{q}, g)$ represents the PDF of parton
$i$ in proton $P(=A,B)$, $\mu_{f}$ is the factorization scale, $x_A$
and $x_B$ describe the momentum fractions of parton (quark or gluon)
in protons $A$ and $B$, respectively. The expression for the total
cross section for $e^+ e^- \to W^+W^-\gamma/Z$ is
\begin{eqnarray}\label{int-eevvv}
\sigma_{ee}^{\gamma/Z} &=& \frac{1}{4|\vec{p}|\sqrt{s}}\int {\rm
d}\Gamma_3 \sum_{spin}{}^\prime|{\cal M}_{ee}^{\gamma/Z}|^2,
\end{eqnarray}
where $\vec{p}$ is the three-momentum of the incoming $e^+$ (or $e^-$) in
the c.m.s of $e^+e^-$ collider. The prime on the
sum means averaging over initial spin states as declared for Eq.(\ref{int-ppvvv}).

\vskip 5mm
\section{Numerical results and discussions}
\par
In this section, we present the numerical results in both the SM and
the LED model at the LHC and ILC.
For the calculations at the LHC, we use the CTEQ6L1 PDFs \cite{19}
with $\Lambda_{QCD} = 165~MeV$ and $n_{f}=5$; the factorization
scale is set to be $\mu_f=m_W$ and $\mu_f=m_W + m_Z/2$ for the $pp
\to W^+ W^- \gamma$ and $pp \to W^+ W^- Z$ processes, respectively.
The active quarks are taken as massless, i.e.,
$m_{q}=0,~(q=u,d,c,s,b)$, the CKM matrix is set to be the unit matrix. The
other related input parameters are taken as \cite{20}
\begin{eqnarray}
~~~~~~\alpha^{-1}(0)=137.036,~~m_W=80.385~\textrm{GeV},~~m_Z=91.1876~\textrm{GeV}, \nb \\
M_{H}=125~\textrm{GeV},~~m_{t}=173.5~\textrm{GeV},~~m_{e}=0.511 \times 10^{-3}~\textrm{GeV}.
\end{eqnarray}
Since the LED model is an effective low-energy theory, it breaks
down in the non-perturbative region where $\sqrt{s}(\sqrt{\hat{s}})
\simeq M_S$ or above. In order to make reliable and viable phenomenological
predictions, we take the hard and conservative truncation scheme as
setting the cut $\sqrt{\hat{s}}<M_S$ for proton-proton collision and
the limit $\sqrt{s}\sim 1$~TeV~$<M_S$ for the $e^+e^-$ collision, where
$\sqrt{\hat{s}}$ and $\sqrt{s}$ are the partonic and $e^+$-$e^-$
c.m.s energies, respectively. In evaluating the $e^+e^-/pp \to
W^+W^-\gamma$ processes, we put a transverse momentum cut
$p_{T}^{\gamma}>25$~GeV and a rapidity cut $|\eta^{\gamma}|<2.7$ on
the final photon in order to get rid of the IR singularity at the
tree level.

\par
Recently, the LED parameters $M_{S}$ and $n$ have obtained more
severe constraints by the LHC experiments. The ATLAS Collaboration provided $95\%$ confidence level
lower limits on $M_{S}$ in the range of $2.27-3.53$~TeV depending on
the number of extra dimensions $n$ in the range of $7$ to $3$ \cite{21}. The
diphoton searches at CMS set $2.3$~TeV~$<M_{S}<3.8$~TeV \cite{22}, and
the dilepton experiments at CMS set the limit on $M_{S}$ as $2.5$~TeV~$<
M_{S} <3.8$~TeV with the number of extra dimensions $n$ varying from $7$ to $3$ at
$95\%$ confidence level \cite{23}. In our calculations we take $M_{S}=3.8$~TeV and
$n=3$ unless otherwise stated.

\par
For the verification of the correctness of our numerical calculations, we
use both the FeynArts 3.5 \cite{17} and CompHEP 4.5.1 \cite{24} packages to
calculate the integrated cross sections for the $e^+e^- \to W^+W^-\gamma/Z$
and $pp \to W^+W^-\gamma/Z$ processes at the
$\sqrt{s}=800$~GeV ILC and the $\sqrt{s}=14$~TeV LHC in the SM separately.
We take the input parameters, PDFs, and the event selection criteria as mentioned
above. The numerical results are listed in Table \ref{tab1}.
It demonstrates that the results from the two packages are in good agreement within the
calculation errors.
\begin{table}[htbp]
\begin{center}
\begin{tabular}{|c|c|c|}
\hline
{} ILC                         &FeynArts[fb]            &CompHEP[fb]  \\
\hline  $e^+e^- \to W^+W^-\gamma$     &99.15(3)      &99.16(2)        \\
\hline  $e^+e^- \to W^+W^-Z$          &52.16(2)      &52.15(1)        \\
\hline
{} LHC                         &FeynArts[fb]       &CompHEP[fb]     \\
\hline  $pp \to W^+W^-\gamma$         &122.84(2)     &122.86(3)       \\
\hline  $pp \to W^+W^-Z$              &90.35(1)      &90.35(1)        \\
\hline
\end{tabular}
\caption{  \label{tab1} The integrated cross sections for the processes $e^+e^-
\to W^+W^-\gamma/Z$ and $pp \to W^+W^-\gamma/Z$ in the SM at the
$\sqrt{s}=800$~GeV ILC and the $\sqrt{s}=14$~TeV LHC by using FeynArts 3.5 \cite{17} and
CompHEP 4.5.1 \cite{24} packages separately. }
\end{center}
\end{table}

\par
\begin{figure}[htbp]
\begin{center}
\includegraphics[scale=0.7]{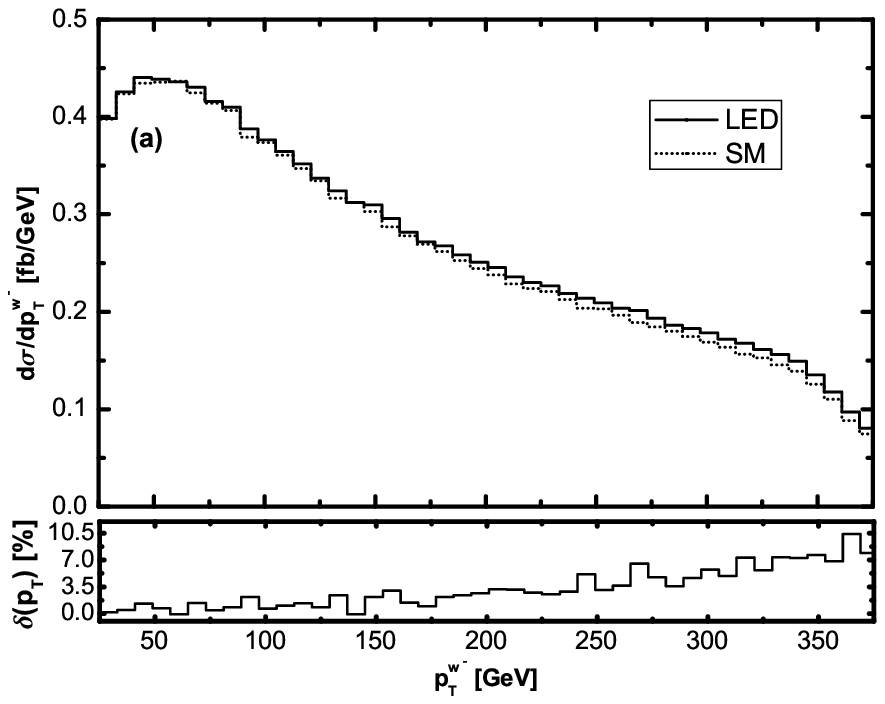}%
\hspace{0in}%
\includegraphics[scale=0.7]{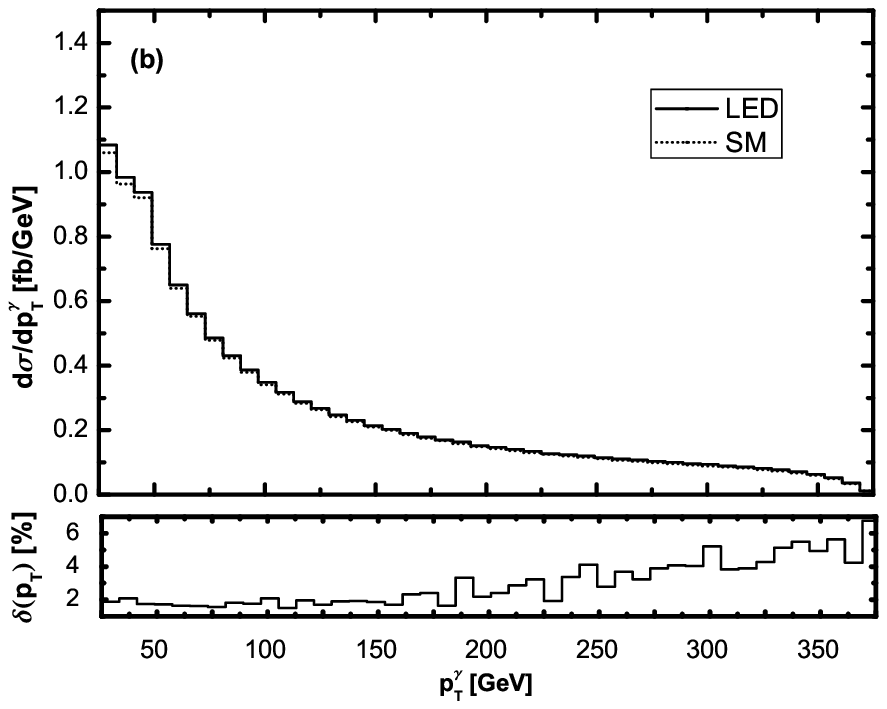}%
\hspace{0in}%
\caption{ \label{fig4} Transverse momentum and corresponding
relative LED discrepancy distributions of final $W^-$ and $\gamma$
for the $e^+e^- \to W^{+}W^{-}\gamma$ process in both the SM and the
LED model at the $\sqrt{s}=800$~GeV ILC, with the LED parameters
$M_S=3.8$~TeV and $n=3$. (a) for $p_T^{W^-}$. (b) for
$p_T^{\gamma}$. }
\end{center}
\end{figure}
\begin{figure}[htbp]
\begin{center}
\includegraphics[scale=0.7]{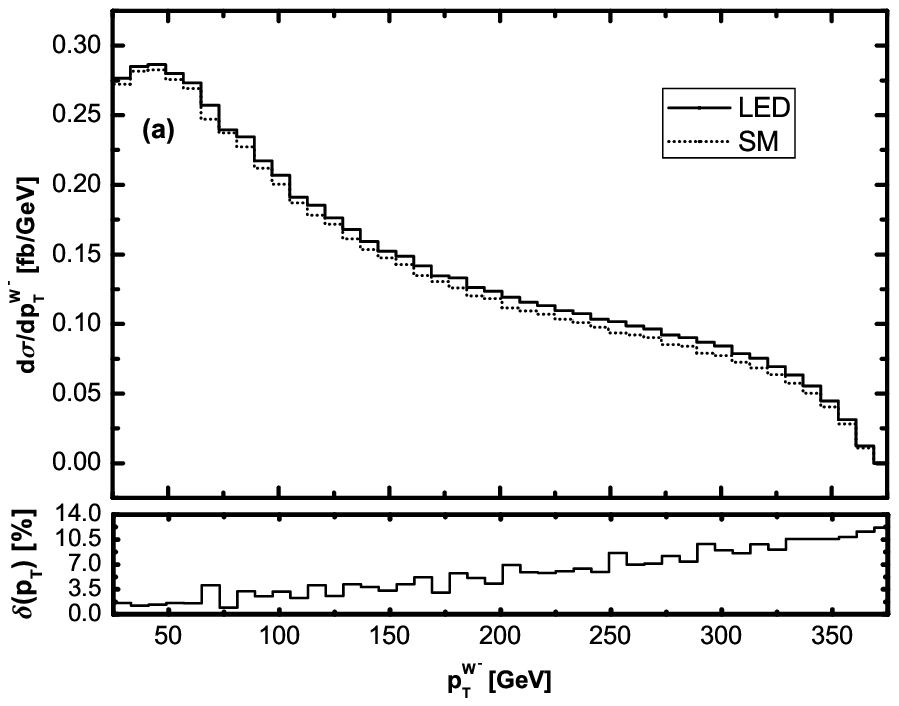}%
\hspace{0in}%
\includegraphics[scale=0.7]{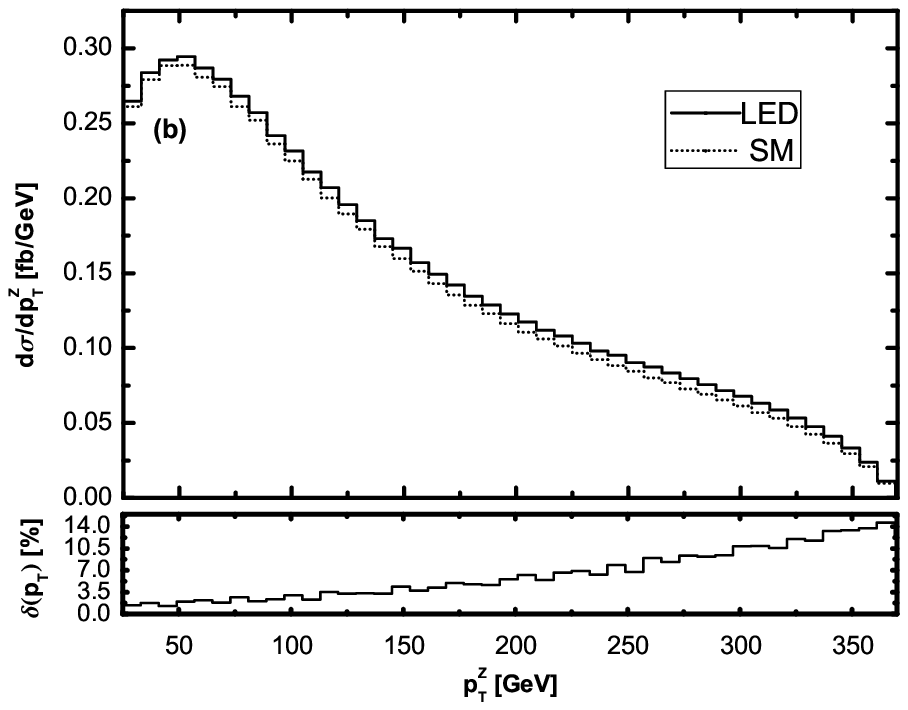}%
\hspace{0in}%
\caption{ \label{fig5}  Transverse momentum and corresponding
relative LED discrepancy distributions of final $W^-$ and $Z$ boson
for the $e^+e^- \to W^{+}W^{-}Z$ process in both the SM and the LED
model at the $\sqrt{s}=800$~GeV ILC, with the LED parameters
$M_S=3.8$~TeV and $n=3$. (a) for $p_T^{W^-}$. (b) for $p_T^{Z}$. }
\end{center}
\end{figure}
\begin{figure}[htbp]
\begin{center}
\includegraphics[scale=0.7]{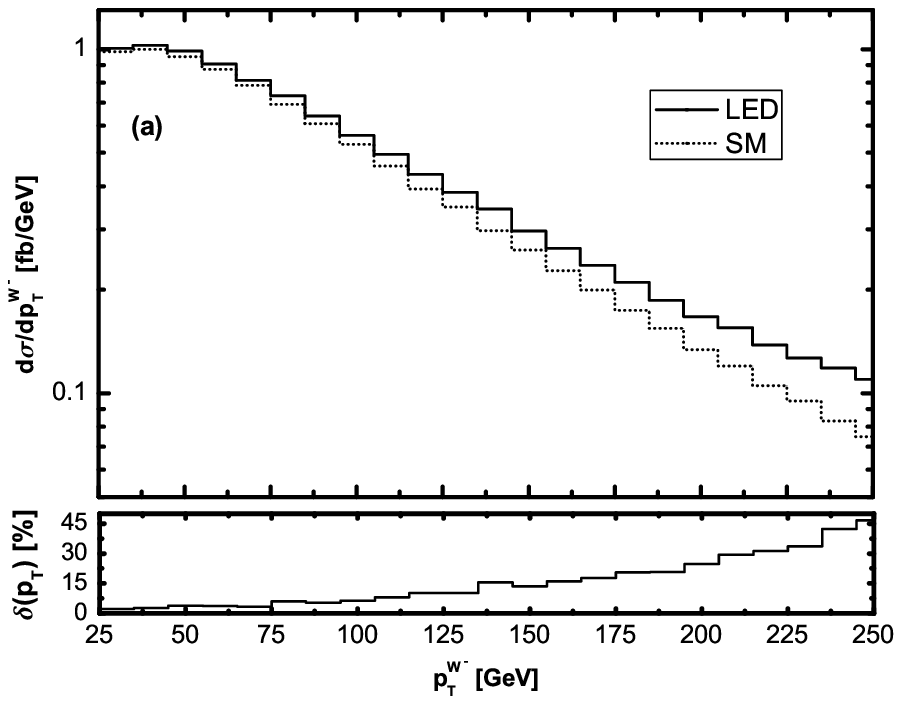}%
\hspace{0in}%
\includegraphics[scale=0.7]{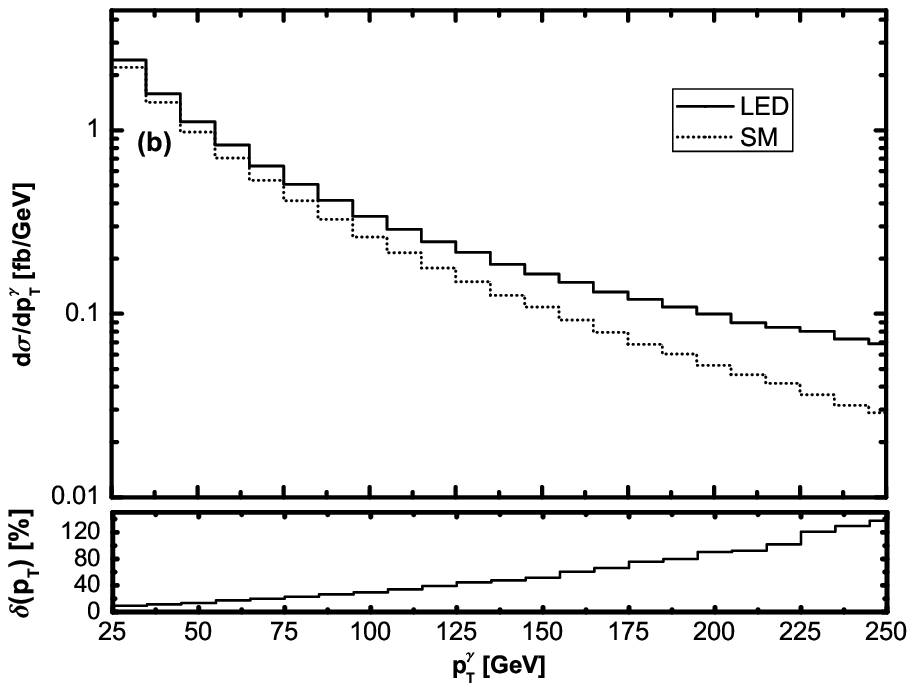}%
\hspace{0in}%
\caption{ \label{fig6} Transverse momentum and corresponding
relative LED discrepancy distributions of final $W^-$ and $\gamma$
for the $pp \to W^{+}W^{-}\gamma$ process in both the SM and the LED
model at the $\sqrt{s}=14$~TeV LHC, with the LED parameters
$M_S=3.8$~TeV and $n=3$. (a) for $p_T^{W^-}$. (b) for
$p_T^{\gamma}$. }
\end{center}
\end{figure}
\begin{figure}[htbp]
\begin{center}
\includegraphics[scale=0.7]{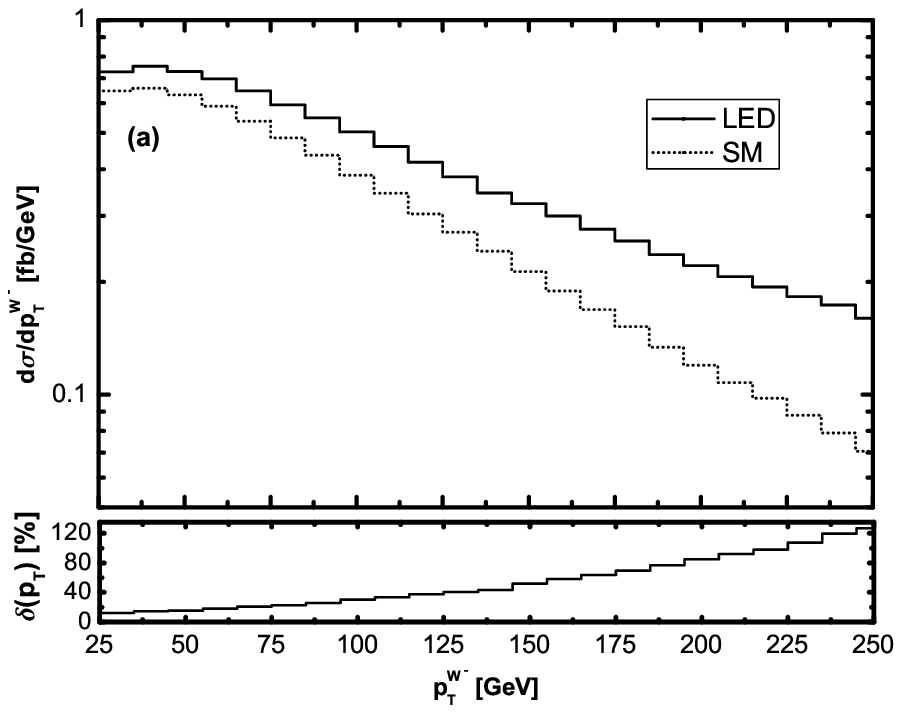}%
\hspace{0in}%
\includegraphics[scale=0.7]{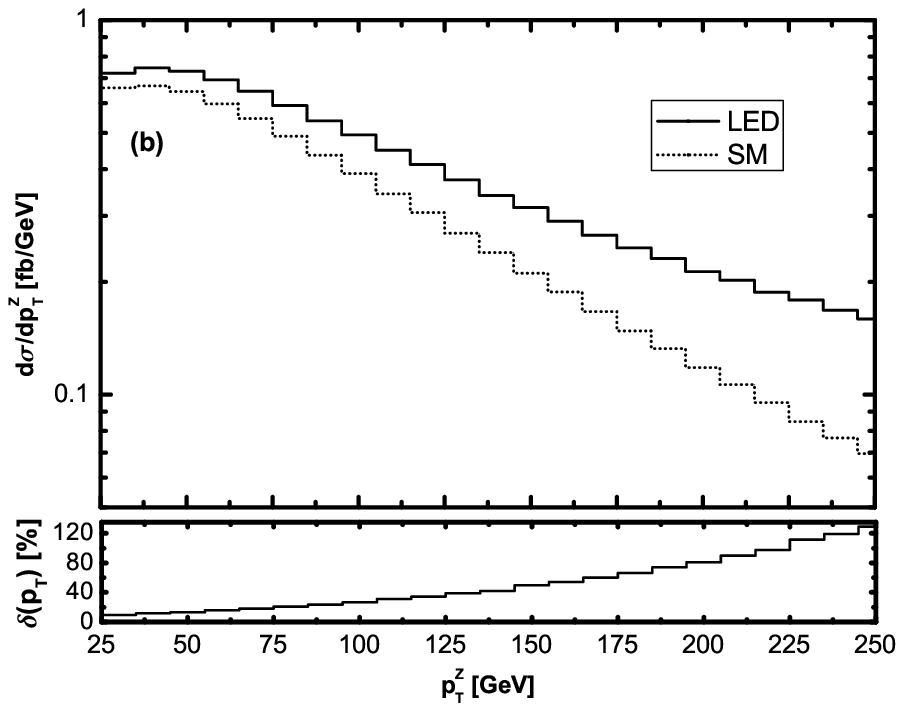}%
\hspace{0in}%
\caption{ \label{fig7} Transverse momentum and corresponding
relative LED discrepancy distributions of final $W^-$ and $Z$ boson
for the $pp \to W^{+}W^{-}Z$ process in both the SM and the LED
model at the $\sqrt{s}=14$~TeV LHC, with the LED parameters
$M_S=3.8$~TeV and $n=3$. (a) for $p_T^{W^-}$. (b) for $p_T^{Z}$. }
\end{center}
\end{figure}

We present the transverse momentum ($p_T$) distributions of final
$W^{-}$ boson, $\gamma$ and $Z$ boson for the $e^+e^- \to
W^+W^-\gamma, W^+W^-Z$ processes at the $\sqrt{s}=800$~GeV ILC in
Figs.\ref{fig4} and \ref{fig5}. The $p_T^{W^-}$, $p_T^Z$
and $p_T^{\gamma}$ distributions for the $pp \to W^+W^-\gamma, W^+W^-Z$
processes at the $\sqrt{s}=14$~TeV LHC are presented in
Figs.\ref{fig6} and \ref{fig7}. In each plot of
Figs.\ref{fig4}-\ref{fig7}, the $p_T^{W^-}$, $p_T^{Z}$ and
$p_T^{\gamma}$ distributions are provided in both the SM and the LED
model for the sake of comparison. We define the relative LED
discrepancy of $p_T$ distribution as $\delta(p_T)\equiv
\left(\frac{d\sigma_{LED}}{dp_T}-
\frac{d\sigma_{SM}}{dp_T}\right)/\frac{d\sigma_{SM}}{dp_T}$ to
describe the LED effect on the differential cross section, and plot the
corresponding $\delta(p_T)$ distribution in the nether plot for each
of the figures in Figs.\ref{fig4}-\ref{fig7}. All 8 figures show
that $\delta(p_T^{W^-})$, $\delta(p_T^{\gamma})$ and
$\delta(p_T^{Z})$ at the ILC and LHC become larger with the
increment of the transverse momenta. Specifically, in
Fig.\ref{fig4}(a) the $\delta(p_T^{W^-})$ lies in the range of
$0.7\%-7.9\%$ for the $e^+e^- \to W^+W^-\gamma$ process in the region of
$25$~GeV~$<p_T^{W^-}<375$~GeV, while $\delta(p_T^{W^-})$ varies from
$1.7\%$ to $12.0\%$ for the $e^+e^- \to W^+W^-Z$ process in the same
$p_T$ region as shown in Fig.5(a). From Figs.\ref{fig4} and
\ref{fig5}, one can find that all the curve behaviors of
$\delta(p_T^{\gamma})$, $\delta(p_T^{Z})$ and $\delta(p_T^{W^-})$
are similar in both the $e^+e^- \to W^+W^-\gamma$ and the $e^+e^-
\to W^+W^-Z$ processes at the ILC. Figs.\ref{fig6} and
\ref{fig7} are for the $pp \to W^+W^-\gamma$ and $pp \to
W^+W^-Z$ processes at the LHC, respectively. We can see that the
LED effects on the $p_T^{W^-}$, $p_T^{\gamma}$ and $p_T^{Z}$
distributions at the LHC become dominant over the pure SM
contributions in the high $p_T$ region. The feature of the $p_T$
distributions at the LHC can serve as LED signal searches in the TGB
measurements.

\par
\begin{figure}[htbp]
\begin{center}
\includegraphics[scale=0.7]{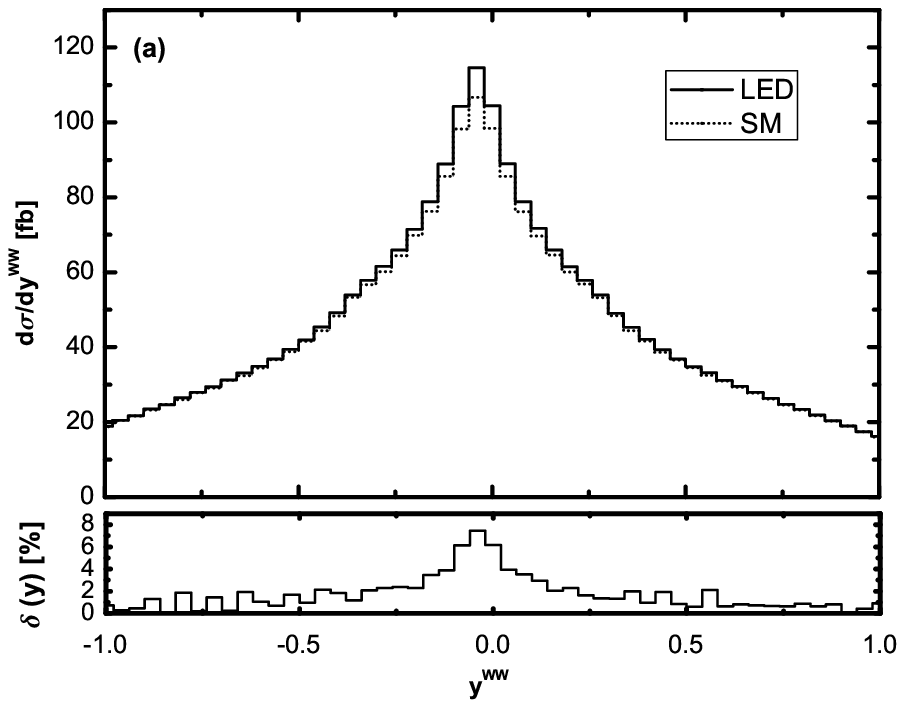}%
\hspace{0in}%
\includegraphics[scale=0.7]{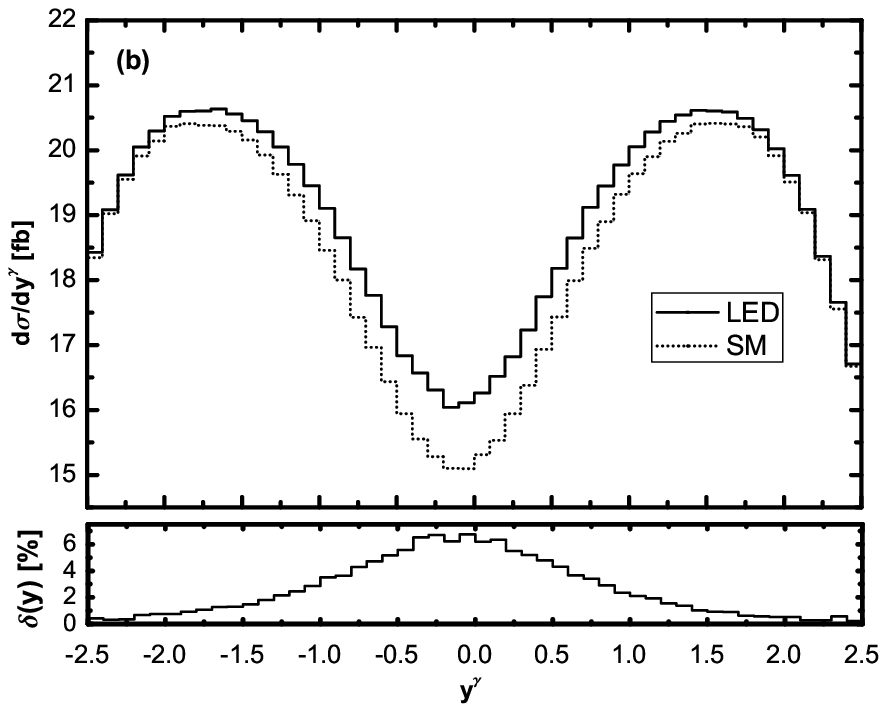}%
\hspace{0in}%
\caption{ \label{fig8} Rapidity and corresponding relative LED
discrepancy distributions of final $W$ pair and $\gamma$ for the
$e^+e^- \to W^{+}W^{-}\gamma$ process in both the SM and the LED
model at the $\sqrt{s}=800$~GeV ILC, with the LED parameters
$M_S=3.8$~TeV and $n=3$. (a) for $y^{WW}$. (b) for $y^{\gamma}$. }
\end{center}
\end{figure}
\begin{figure}[htbp]
\begin{center}
\includegraphics[scale=0.7]{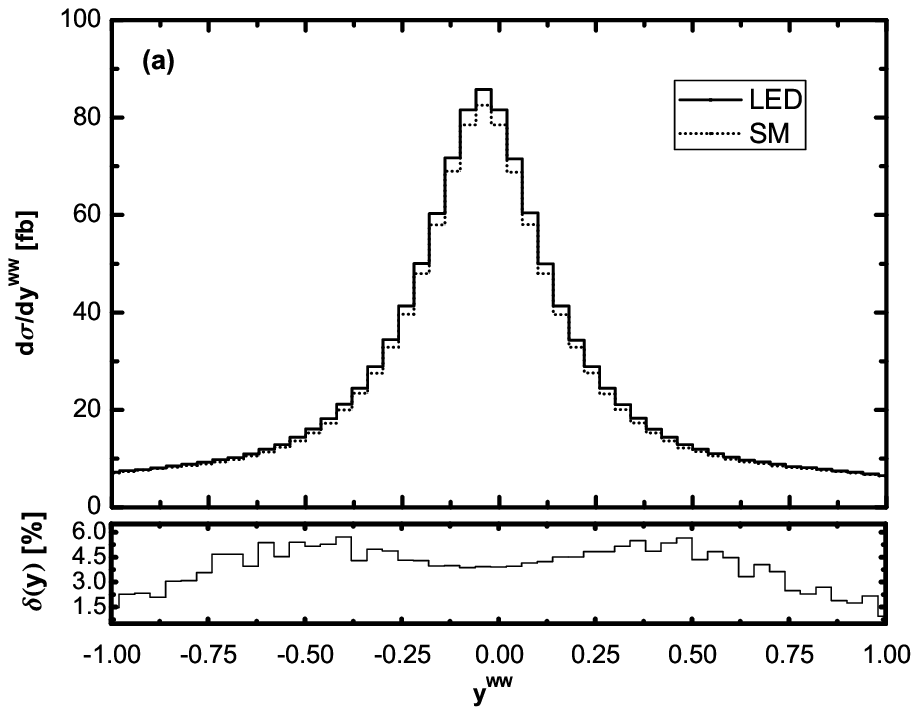}%
\hspace{0in}%
\includegraphics[scale=0.7]{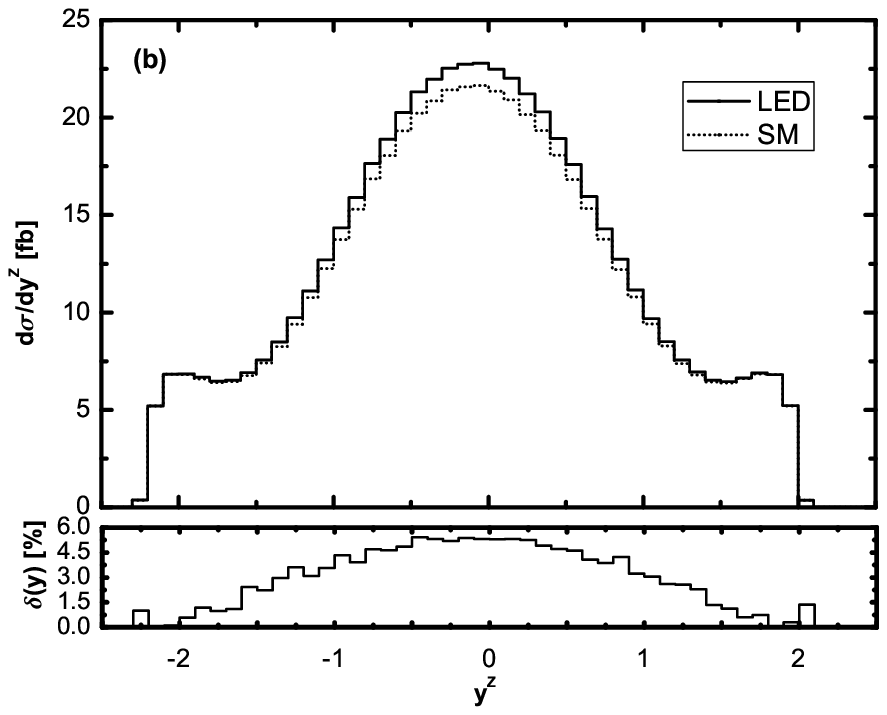}%
\hspace{0in}%
\caption{ \label{fig9} Rapidity and corresponding relative LED
discrepancy distributions of final $W$ pair and $Z$ boson for the
$e^+e^- \to W^{+}W^{-}Z$ process in both the SM and the LED model at
the $\sqrt{s}=800$~GeV ILC, with the LED parameters $M_S=3.8$~TeV
and $n=3$. (a) for $y^{WW}$. (b) for $y^{Z}$. }
\end{center}
\end{figure}
\begin{figure}[htbp]
\begin{center}
\includegraphics[scale=0.7]{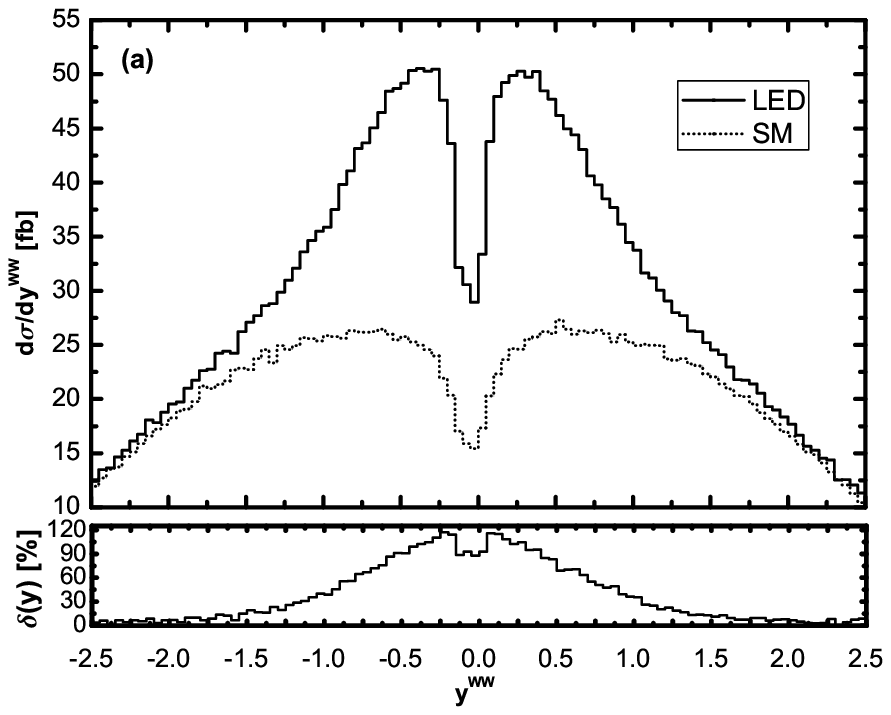}%
\hspace{0in}%
\includegraphics[scale=0.7]{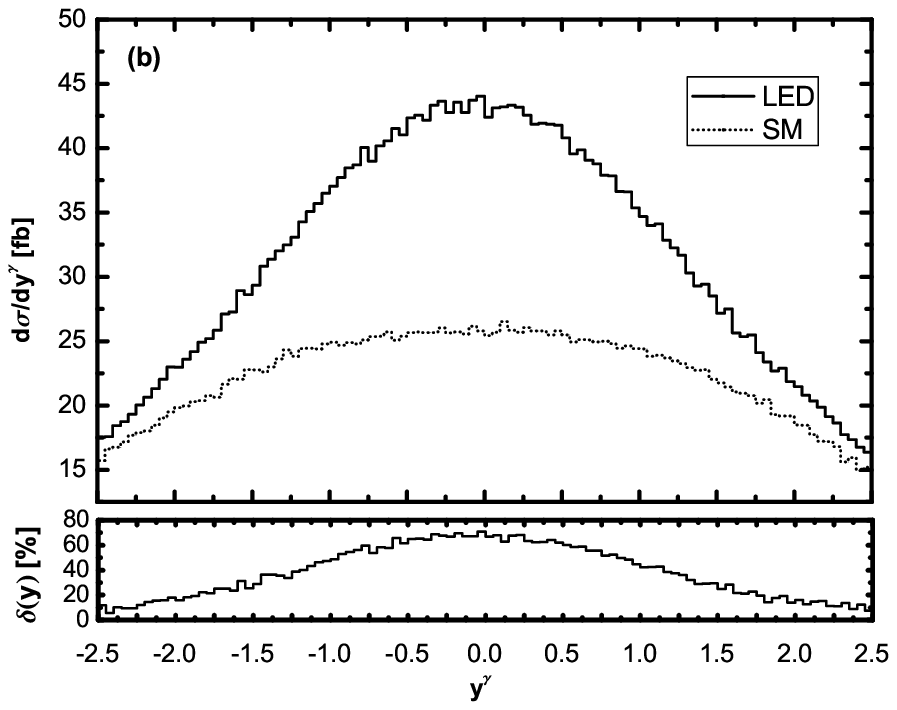}%
\hspace{0in}%
\caption{ \label{fig10} Rapidity and corresponding relative LED
discrepancy distributions of final $W$ pair and $\gamma$ for the $pp
\to W^{+}W^{-}\gamma$ process in both the SM and the LED model at
the $\sqrt{s}=14$~TeV LHC, with the LED parameters $M_S=3.8$~TeV and
$n=3$. (a) for $y^{WW}$. (b) for $y^{\gamma}$.  }
\end{center}
\end{figure}
\begin{figure}[htbp]
\begin{center}
\includegraphics[scale=0.7]{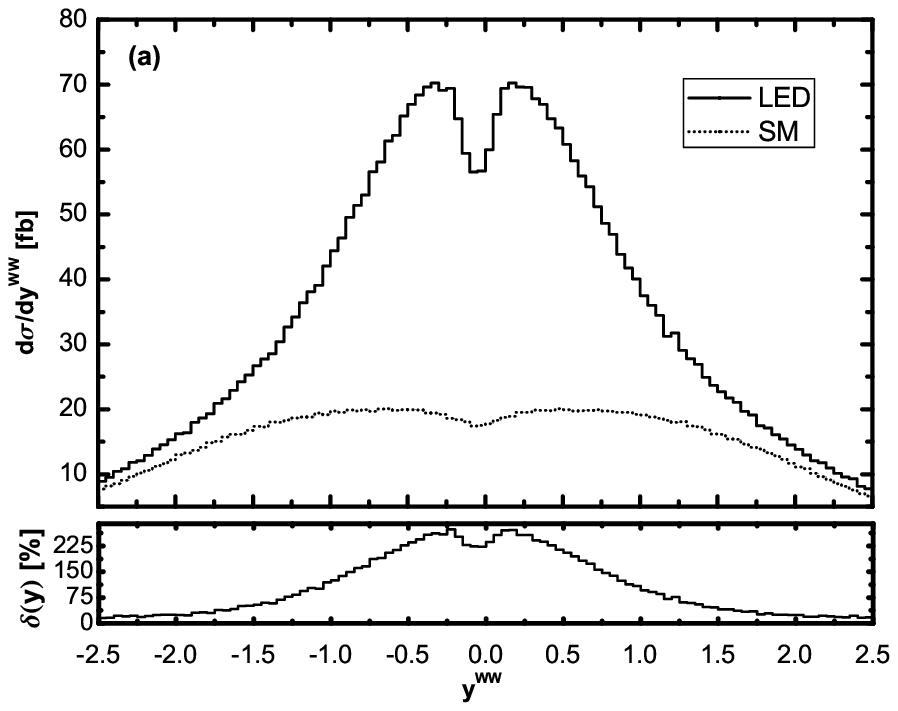}%
\hspace{0in}%
\includegraphics[scale=0.7]{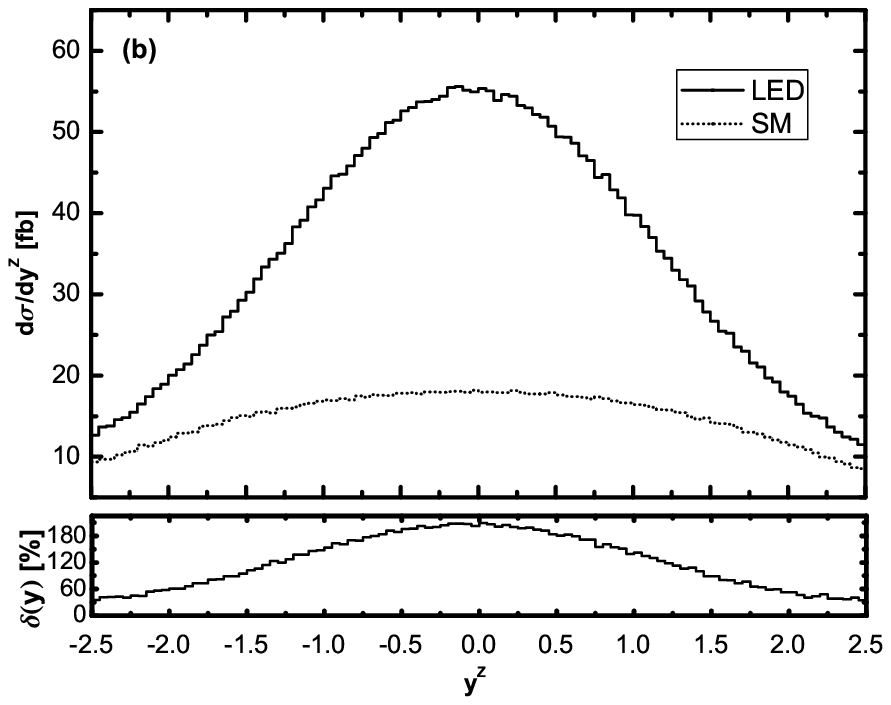}%
\hspace{0in}%
\caption{ \label{fig11}  Rapidity and corresponding relative LED
discrepancy distributions of final $W$ pair and $Z$ boson for the
$pp \to  W^{+}W^{-}Z$ process in both the SM and the LED model at
the $\sqrt{s}=14$~TeV LHC, with the LED parameters $M_S=3.8$~TeV and
$n=3$. (a) for $y^{WW}$. (b) for $y^{Z}$.  }
\end{center}
\end{figure}

In Figs.\ref{fig8} and Figs.\ref{fig9}, we depict the rapidity ($y$)
distributions of final $W$ pair, $\gamma$ and $Z$ boson for the
$e^+e^- \to W^+W^-\gamma, W^+W^-Z$ processes at the
$\sqrt{s}=800$~GeV ILC, respectively. Figure.\ref{fig10} show the
$y^{WW}$ and $y^{\gamma}$ distributions of the process $pp \to
W^+W^-\gamma$, while Fig.\ref{fig11} give the $y^{WW}$ and $y^{Z}$
distributions for the process $pp \to W^+W^-Z$. In each plot of
Figs.\ref{fig8}-\ref{fig11}, there are $y^{WW}$ and
$y^{\gamma}(y^{Z})$ rapidity distributions in both the SM and
the LED model, and the corresponding relative LED discrepancy
$\delta(y)$ distribution, where we define $\delta(y)\equiv
\left(\frac{d\sigma_{LED}}{dy}- \frac{d\sigma_{SM}}{dy}\right)
/\frac{d\sigma_{SM}}{dy}$. We can see from all the figures that the
contributions from the LED manifest themselves obviously in the
central rapidity regions at both colliders. From
Figs.\ref{fig8}(b), \ref{fig9}(b),\ref{fig10}(b) and
Fig.\ref{fig11}(b), we see that the relative LED discrepancies,
$\delta(y^{\gamma})$ and $\delta(y^{Z})$, respectively reach their
peaks at the locations of $y^{\gamma(Z)}\sim 0$ with maximum values
about $6\%$ at the ILC and beyond $50\%$ ($140\%$) for $y^{\gamma}$
($y^{Z}$) at the LHC.

\par
\begin{figure}[htbp]
\begin{center}
\includegraphics[scale=0.7]{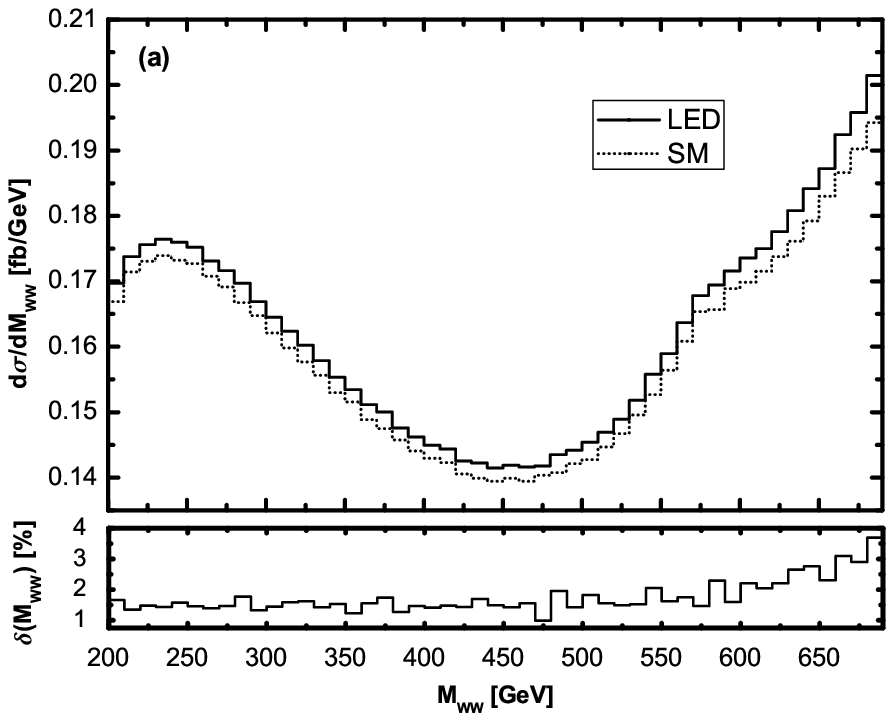}%
\hspace{0in}%
\includegraphics[scale=0.7]{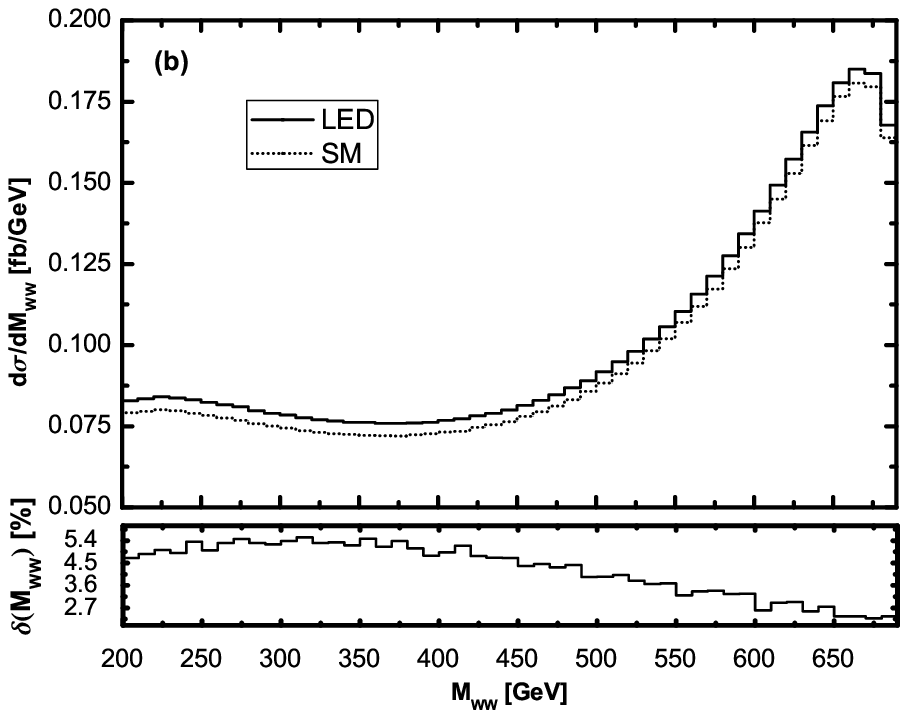}%
\hspace{0in}%
\caption{ \label{fig12}  $M_{WW}$ and
corresponding relative LED discrepancy distributions in both the SM
and the LED model at the $\sqrt{s}=800$~GeV ILC, with the LED
parameters $M_S=3.8$~TeV and $n=3$. (a) for the $e^+e^- \to
W^{+}W^{-}\gamma$ process. (b) for the $e^+e^- \to W^{+}W^{-}Z$
process. }
\end{center}
\end{figure}
\begin{figure}[htbp]
\begin{center}
\includegraphics[scale=0.7]{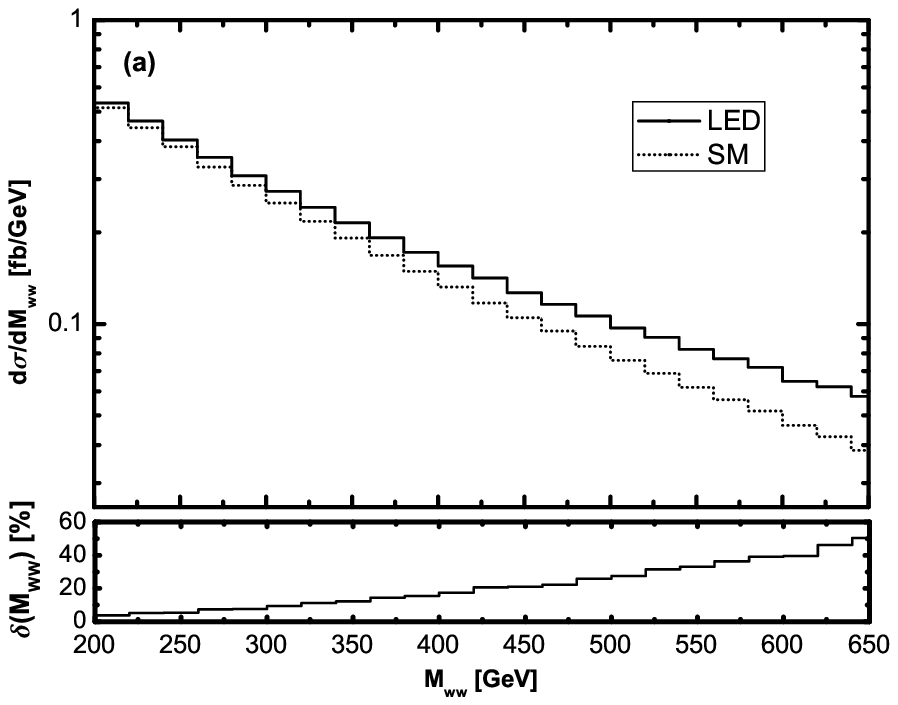}%
\hspace{0in}%
\includegraphics[scale=0.7]{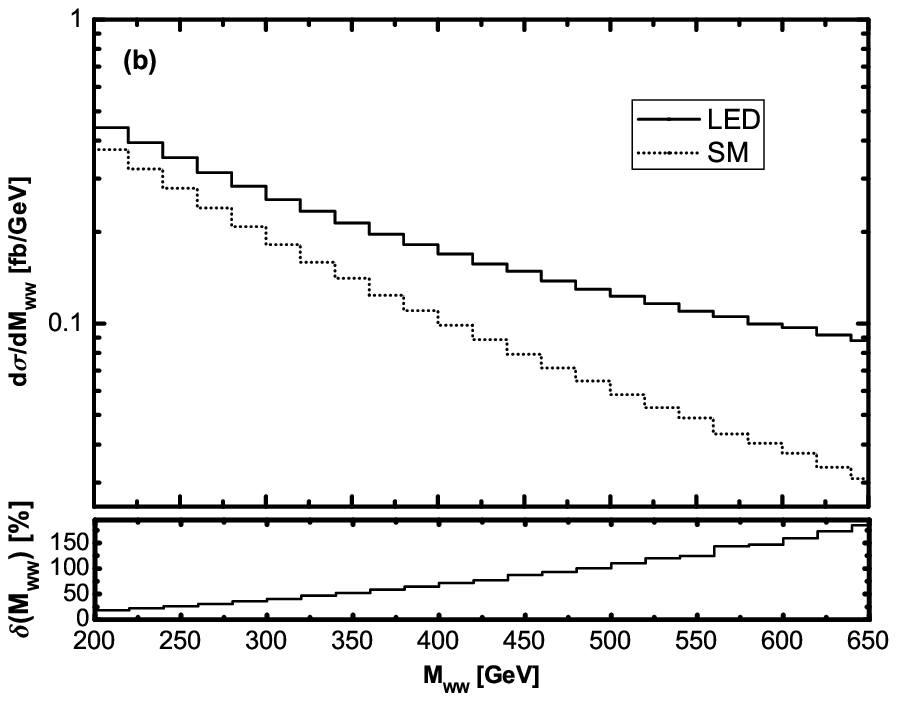}%
\hspace{0in}%
\caption{ \label{fig13}  $M_{WW}$ and
corresponding relative LED discrepancy distributions in both the SM
and the LED model at the $\sqrt{s}=14$~TeV ILC, with the LED
parameters $M_S=3.8$~TeV and $n=3$. (a) for the $pp \to
W^{+}W^{-}\gamma$ process. (b) for the $pp \to W^{+}W^{-}Z$ process.
}
\end{center}
\end{figure}

Figures \ref{fig12} and \ref{fig13} show the $W$-pair invariant mass
($M_{WW}$) distributions in both the SM and the LED model at the ILC
and LHC, respectively. Their corresponding relative LED
discrepancies $\left(\delta(M_{WW})\equiv\left(\frac{d\sigma_{LED}}
{dM_{WW}}- \frac{d\sigma_{SM}}{dM_{WW}}\right)
/\frac{d\sigma_{SM}}{dM_{WW}}\right)$ are also illustrated there.
Figure \ref{fig12}(a) shows that $\delta(M_{WW})$ for the $e^+e^- \to
W^{+}W^{-}\gamma$ process increases when $M_{WW}$ goes up in the plotted range,
and reaches its maximum of about $3.7\%$ at the position of $M_{WW}=690$~GeV.
In contrast, we can see from Fig.\ref{fig12}(b) that $\delta(M_{WW})$ for the $e^+e^-
\to W^{+}W^{-}Z$ process rises gradually until it reaches the maximal value of 
about $5.5\%$ at the postion of $M_{WW}=320$~GeV, and then decreases with the
increment of $M_{WW}$. In Fig.\ref{fig13}, the $M_{WW}$
distributions for the two processes $pp \to W^{+}W^{-}\gamma$ and $pp
\to W^{+}W^{-}Z$ demonstrate the similar behavior; it shows that the
LED effect is going to be dominant with the increment of invariant
mass of the $W$ pair at the LHC.

\par
\begin{figure}[htbp]
\begin{center}
\includegraphics[scale=0.7]{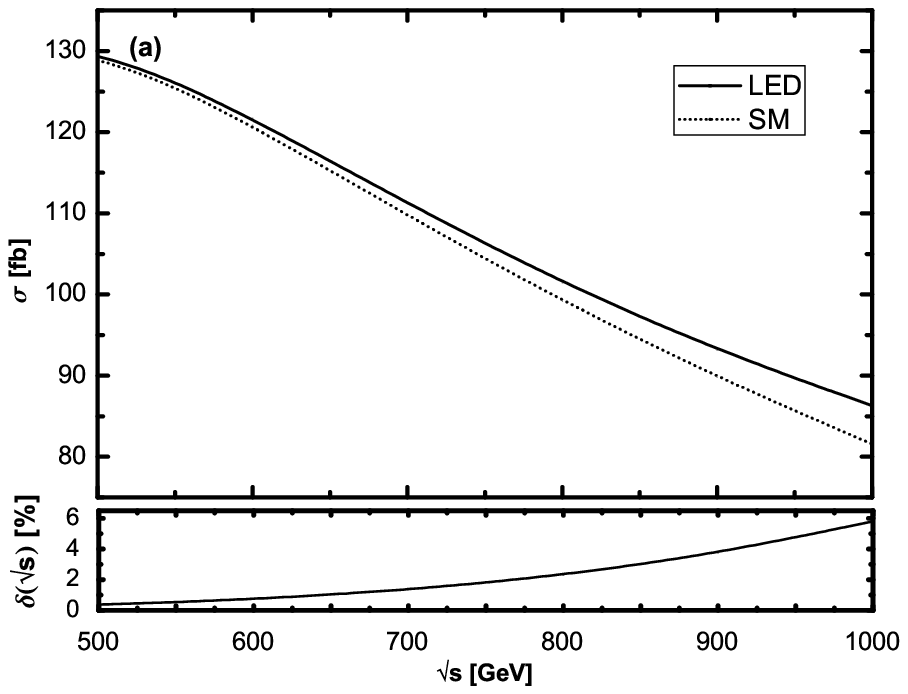}%
\hspace{0in}%
\includegraphics[scale=0.7]{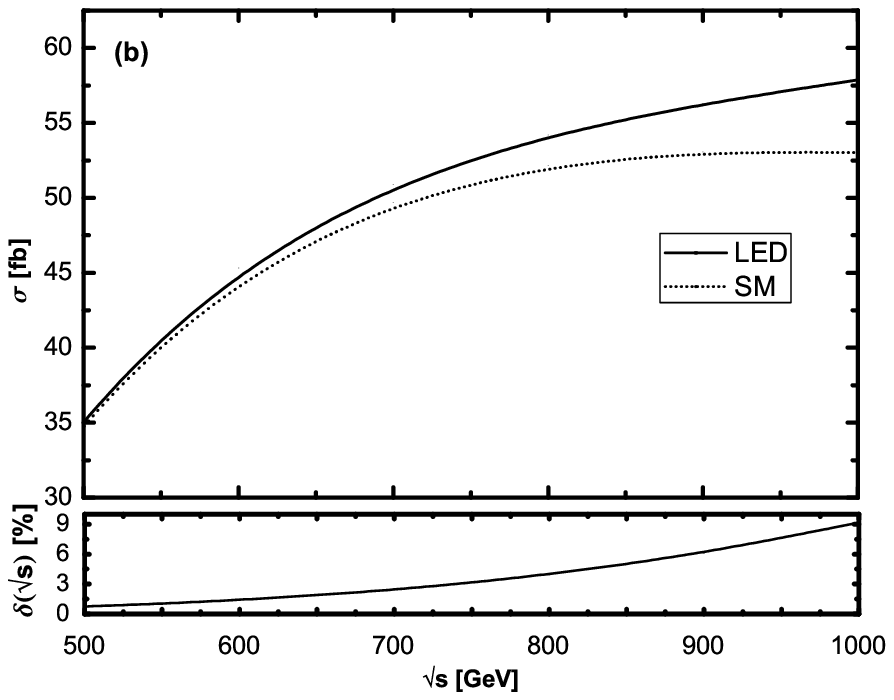}%
\hspace{0in}%
\caption{ \label{fig14} The integrated cross section and the
corresponding relative LED discrepancy as the functions of the c.m.s
energy $\sqrt{s}$ in both the SM and the LED model at the ILC, with
$\sqrt{s}$ varying from $500$~GeV to $1$~TeV, the LED parameters
$M_S=3.8$~TeV and $n=3$. (a) for the $e^+e^- \to W^{+}W^{-}\gamma$
process. (b) for the $e^+e^- \to W^{+}W^{-}Z$ process. }
\end{center}
\end{figure}
\begin{figure}[htbp]
\begin{center}
\includegraphics[scale=0.7]{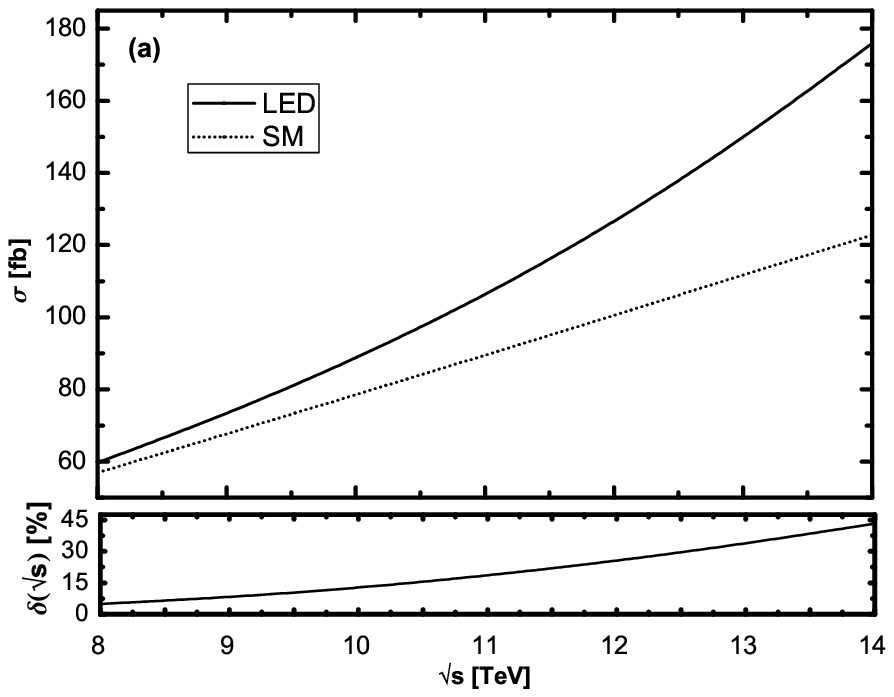}%
\hspace{0in}%
\includegraphics[scale=0.7]{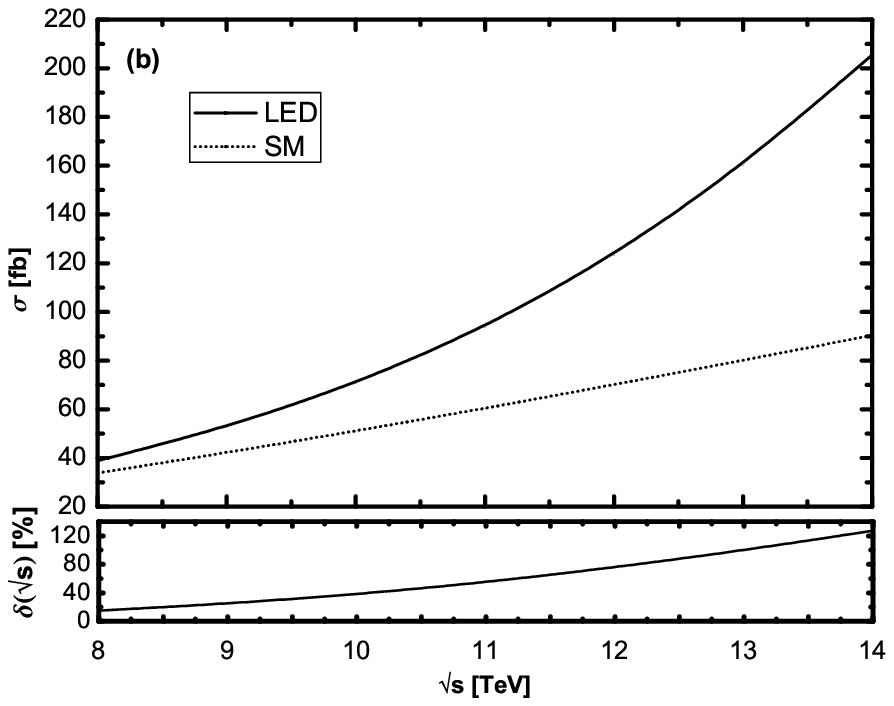}%
\hspace{0in}%
\caption{ \label{fig15} The integrated cross section and the
corresponding relative LED discrepancy as the functions of the c.m.s
energy $\sqrt{s}$ in both the SM and the LED model at the LHC, with
$\sqrt{s}$ varying from $8$ to $14$~TeV, the LED parameters
$M_S=3.8$~TeV and $n=3$. (a) for the $pp \to W^{+}W^{-}\gamma$
process. (b) for the $pp \to W^{+}W^{-}Z$ process.}
\end{center}
\end{figure}

In Figs.\ref{fig14} and \ref{fig15}, we give the integrated cross
section and the corresponding relative LED discrepancy, defined
as $\delta(\sqrt{s})\equiv\frac{\sigma_{LED}-
\sigma_{SM}}{\sigma_{SM}}$, as functions of the c.m.s energy
$\sqrt{s}$ at the ILC and LHC, respectively. From
Fig.\ref{fig14}(a), we see that the distribution of the cross section for
the $e^+e^- \to W^{+}W^{-}\gamma$ process decreases with the
increment of $\sqrt{s}$, while in Fig.\ref{fig14}(b) it shows that the
distribution for the $e^+e^- \to W^{+}W^{-}Z$ process behaves in the
opposite way. In Fig.\ref{fig15}, the cross sections for the $pp
\to W^{+}W^{-}\gamma,W^{+}W^{-}Z$ processes rise sharply when
$\sqrt{s}$ goes up, and the relative LED discrepancies
$\delta(\sqrt{s})$ at the LHC are quantitatively almost one order
larger than those at the ILC.

\par
All the figures above show that there exists a remarkable enhancement of the
relative LED discrepancy at the LHC in comparison with that at
the ILC. This can be ascribed to the following two reasons: (1)
Unlike the $pp \to W^+W^-\gamma/Z$ process in the SM, this process
in the LED model arises from not only the $q\bar q$ annihilations,
but also the additional $gg$ fusion subprocess. In addition, the 
large gluon luminocity at the LHC can further enhance the LED
effect. (2) Since the continuous large colliding energy
$(\sqrt{\hat{s}})$ spectrum available at the LHC can be in the range
not far away from the cutoff scale $\Lambda$ taken to be $M_S$, the resonant
contribution of the single KK-graviton mode in the s-channel will be included.

\par
In Table \ref{tab2} and Table \ref{tab3}, We list the values of the
integrated cross sections in the SM and the LED model for the
$e^+e^- \to W^{+}W^{-}\gamma/Z$ and the $pp \to W^{+}W^{-}\gamma/Z$
processes with some typical colliding energies at the ILC and 
LHC, separately. In Table \ref{tab3}, we give additionally the
cross section contributions from the $pp \to gg \to
W^+W^-\gamma,~W^+W^-Z$ processes. Comparing the results in Table 2
and 3, we can see that in the $pp$ collision case the
existance of the additional $gg$ fusion partonic process and 
the KK-graviton resonance effect induced by the large colliding 
energy spectrum obviously enhance the SM cross section.
\begin{table}[htbp]
\begin{center}
\begin{tabular}{|c|c|c|c|c|}
\hline {}$\sqrt{s}$       &\multicolumn{2}{c|}{$e^+e^- \to
W^+W^-\gamma$}   &\multicolumn{2}{c|}{$e^+e^- \to W^+W^-Z$}\\
\cline{2-5}
{}[GeV]                       &$\sigma_{SM}$[fb]&$\sigma_{LED}$[fb]  &$\sigma_{SM}$[fb]&$\sigma_{LED}$[fb] \\
\hline  500    &130.36(3)     &130.82(3)              &35.70(1)        &35.96(1)                 \\
\hline  600    &120.87(3)     &121.74(3)              &44.71(1)        &45.33(1)                 \\
\hline  700    &109.71(3)     &111.16(3)              &49.70(1)        &50.89(1)                 \\
\hline  800    &99.15(2)      &101.41(3)              &52.16(1)        &54.20(1)                 \\
\hline  900    &89.76(2)      &93.10(2)               &53.07(1)        &56.30(2)                 \\
\hline  1000   &81.58(2)      &86.30(2)               &53.01(1)        &57.86(2)                 \\
\hline
\end{tabular}
\caption{  \label{tab2} The integrated cross sections for the
$e^+e^- \to W^+W^-\gamma$ and $e^+e^- \to W^+W^-Z$ processes in both
the SM and the LED model at the ILC, with the LED parameters
$M_S=3.8$~TeV, $n=3$ and $\sqrt{s}$ varying from $500$~GeV to
$1$~TeV.}
\end{center}
\end{table}
\begin{table}[htbp]
\begin{center}
\begin{tabular}{|c|c|c|c|c|c|c|}
\hline {}$\sqrt{s}$  &\multicolumn{3}{c|}{$pp \to W^+W^-\gamma$}
&\multicolumn{3}{c|}{$pp \to W^+W^-Z$}\\ \cline{2-7} {}[TeV]
&$\sigma_{SM}$[fb]&$\sigma_{LED}(total)$[fb]&$\sigma_{LED}($gg$)$[fb]
&$\sigma_{SM}$[fb]&$\sigma_{LED}(total)$[fb]&$\sigma_{LED}($gg$)$[fb] \\
\hline  8    &56.92(1)    &59.74(1)    &0.45(1)           &33.83(1)    &38.92(1)    &1.27(2)             \\
\hline  9    &67.64(1)    &73.14(1)    &1.06(1)           &44.22(1)    &52.61(1)    &3.00(3)             \\
\hline  10   &78.52(1)    &88.41(1)    &2.21(2)           &51.10(1)    &70.45(1)    &6.28(4)             \\
\hline  11   &89.50(1)    &105.92(1)   &4.20(3)           &60.40(1)    &93.56(1)    &11.91(5)             \\
\hline  12   &100.56(2)   &126.14(2)   &7.30(5)           &70.06(1)    &123.07(1)   &20.81(7)             \\
\hline  13   &111.69(2)   &149.38(2)   &11.87(6)          &80.05(1)    &160.05(2)   &33.94(1)             \\
\hline  14   &122.84(2)   &175.96(2)   &18.25(7)          &90.35(1)    &205.55(3)   &52.35(2)             \\
\hline
\end{tabular}
\caption{  \label{tab3} The integrated cross sections for the $pp
\to W^+W^-\gamma$ and the $pp \to W^+W^-Z$ processes in both the SM
and the LED model at the LHC, with the LED parameters $M_S=3.8$~TeV,
$n=3$ and $\sqrt{s}$ varying from $8$~TeV to $14$~TeV, where the
$\sigma_{LED}({\rm total})$ is the integrated cross section via both
$q$-$\bar{q}$ annihilation and $gg$ fusion, while $\sigma_{LED}(gg)$
denotes the integrated cross section only via $gg$ fusion. }
\end{center}
\end{table}

\par
\begin{figure}[htbp]
\begin{center}
\includegraphics[scale=0.7]{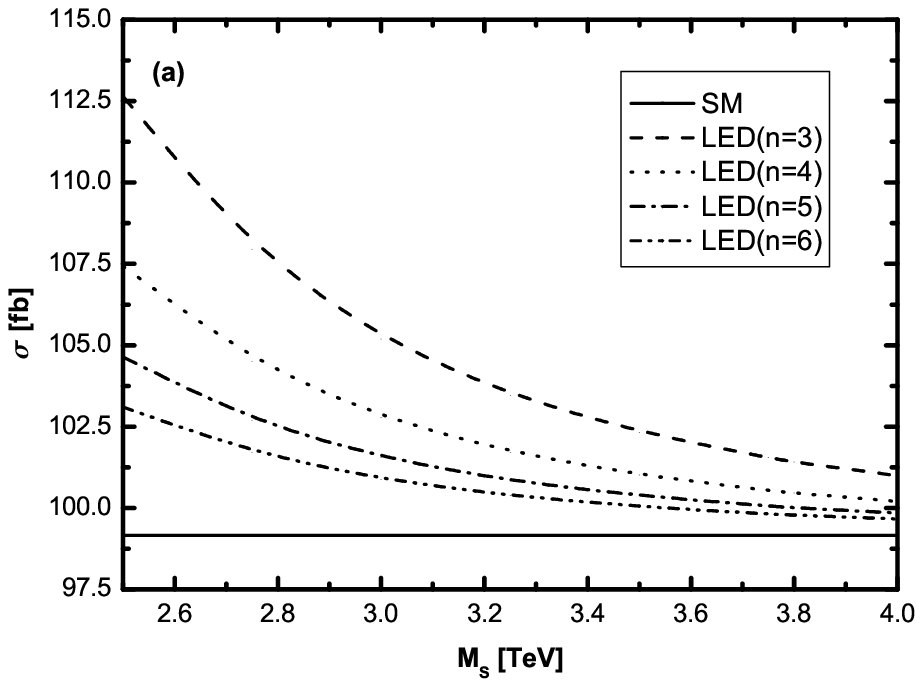}%
\hspace{0in}%
\includegraphics[scale=0.7]{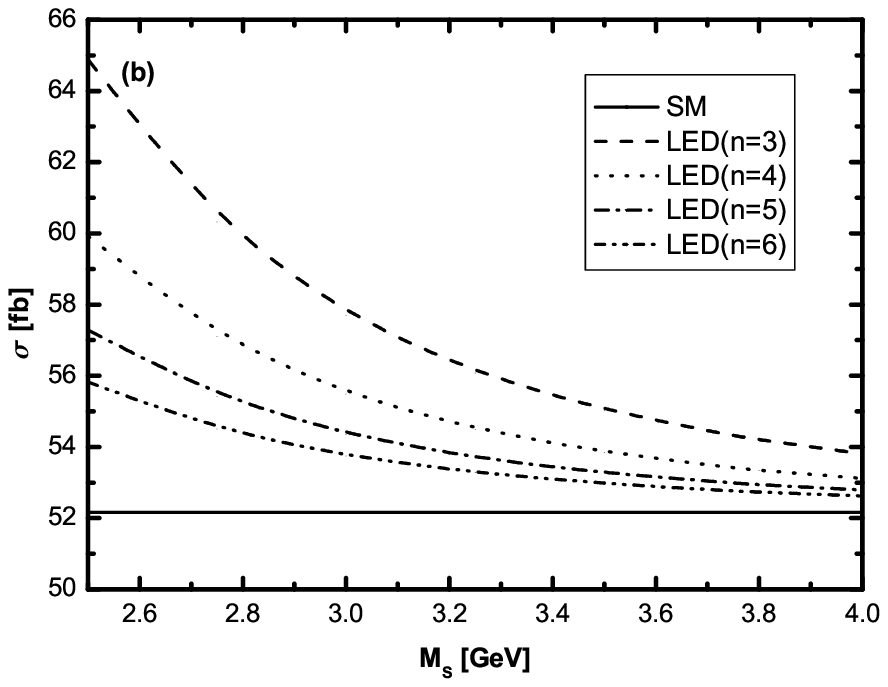}%
\hspace{0in}%
\caption{ \label{fig16} The integrated cross sections as functions of
$M_S$ with $n$ varying from $3$ to $6$ at the $\sqrt{s}=800$~GeV
ILC. The SM results appear as the straight lines. (a) for the
$e^+e^- \to W^{+}W^{-}\gamma$ process. (b) for the $e^+e^- \to
W^{+}W^{-}Z$ process. }
\end{center}
\end{figure}
\begin{figure}[htbp]
\begin{center}
\includegraphics[scale=0.7]{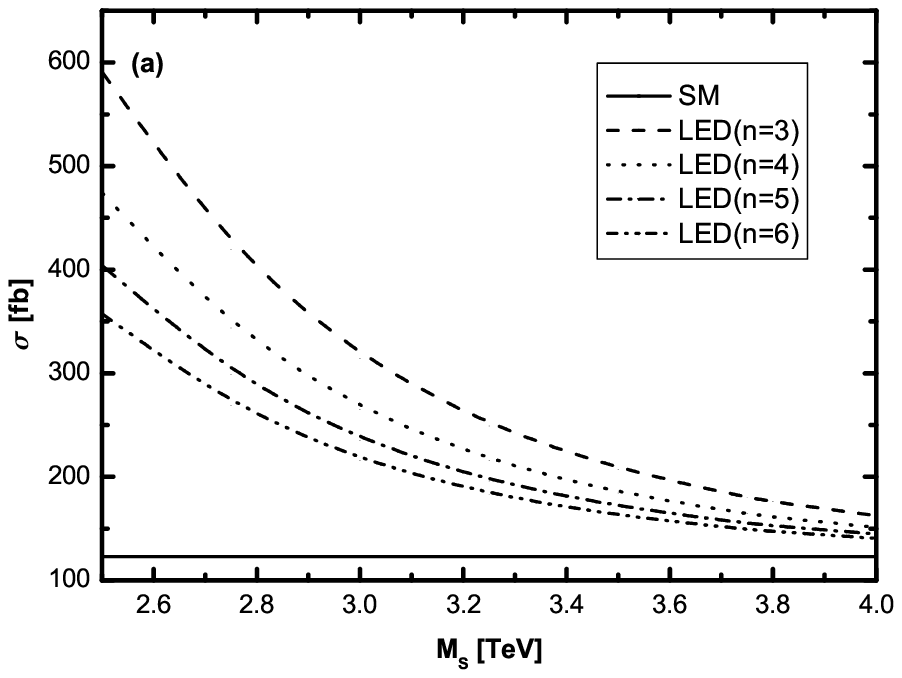}%
\hspace{0in}%
\includegraphics[scale=0.7]{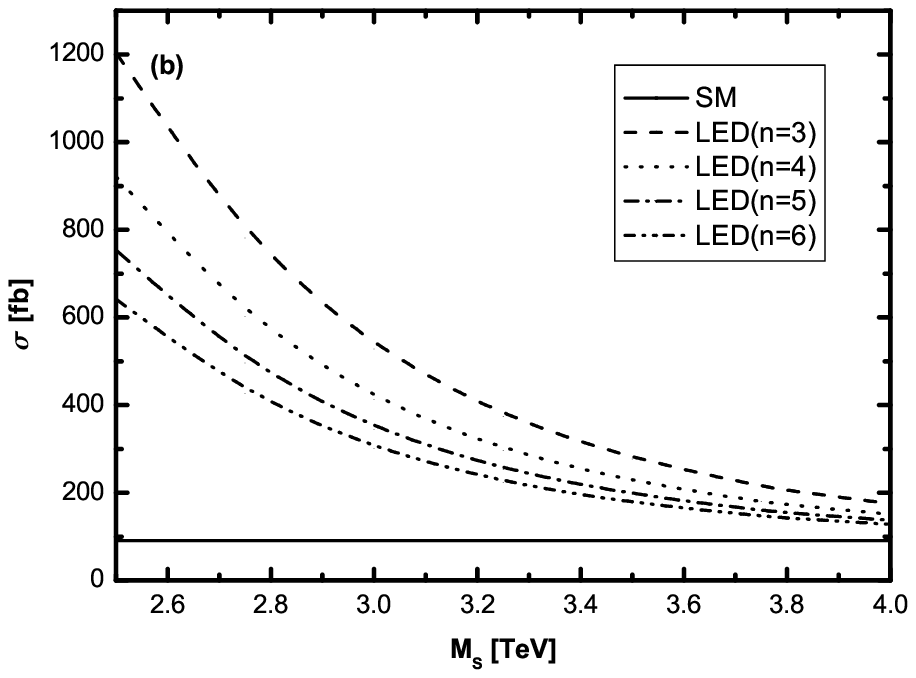}%
\hspace{0in}%
\caption{ \label{fig17} The integrated cross sections as functions of
$M_S$ with $n$ varying from $3$ to $6$ at the $\sqrt{s}=14$~TeV LHC.
The SM results appear as the straight lines. (a) for the $pp \to
W^{+}W^{-}\gamma$ process. (b) for the $pp \to W^{+}W^{-}Z$ process.
}
\end{center}
\end{figure}

In Figs.\ref{fig16} and \ref{fig17}, we present the
integrated cross sections as functions of $M_S$ with different numbers of
extra dimensions $n$ for the $W^{+}W^{-}\gamma/Z$ production
processes at the ILC and LHC, respectively. The horizon line
in each figure corresponds to the SM cross section which is
independent of the LED parameters $M_S$ and $n$. It can be found
that the deviations due to the LED contributions from the SM
predictions become more distinct when $n$ is small. On the other
hand, for a fixed value of $n$, the integrated cross section decreases
with the increment of $M_S$ and gradually approaches to the SM
prediction. These features of the relationship between the integrated
cross section and the LED parameters $M_S$ and $n$ are manifested in
the $W^{+}W^{-}\gamma/Z$ production processes at both the ILC and
LHC, which can be seen in Figs.\ref{fig16} and
\ref{fig17}.

\par
\begin{figure}[htbp]
\begin{center}
\includegraphics[scale=0.7]{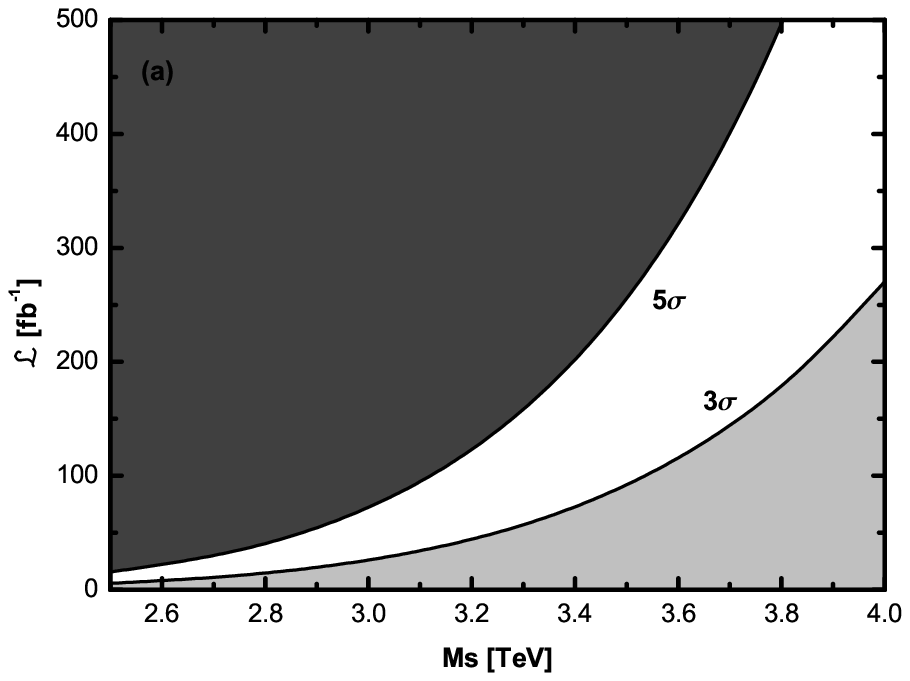}%
\hspace{0in}%
\includegraphics[scale=0.7]{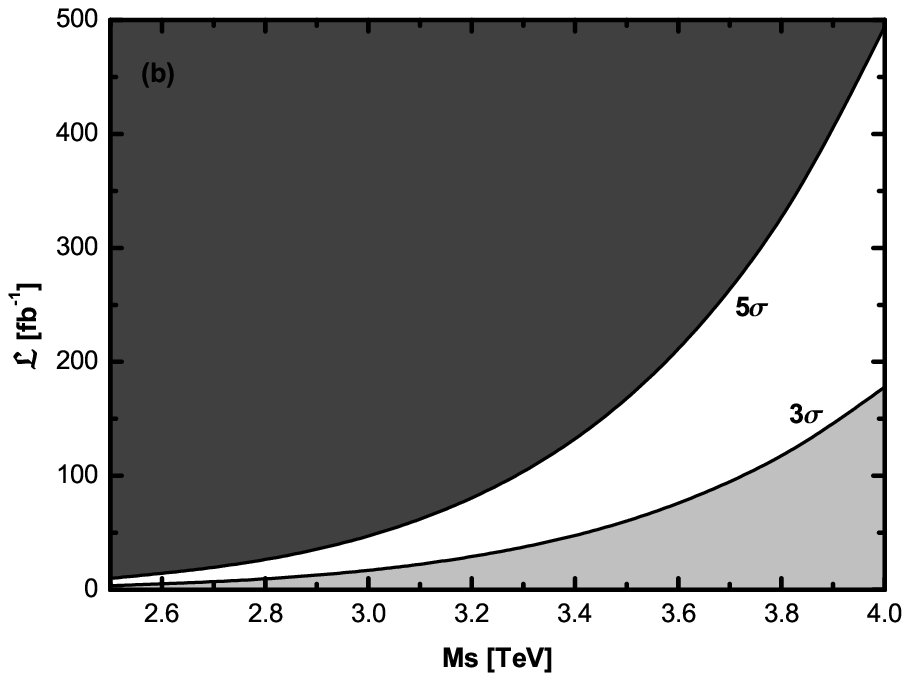}%
\hspace{0in}%
\caption{ \label{fig18} The LED effect discovery area (dark) and
exclusion area (gray) in the $\mathcal{L}-M_S$ space at the
$\sqrt{s}=800$~GeV ILC, where $n=3$. (a) for the $e^+e^- \to
W^{+}W^{-}\gamma$ process. (b) for the $e^+e^- \to W^{+}W^{-}Z$
process. }
\end{center}
\end{figure}
\begin{figure}[htbp]
\begin{center}
\includegraphics[scale=0.7]{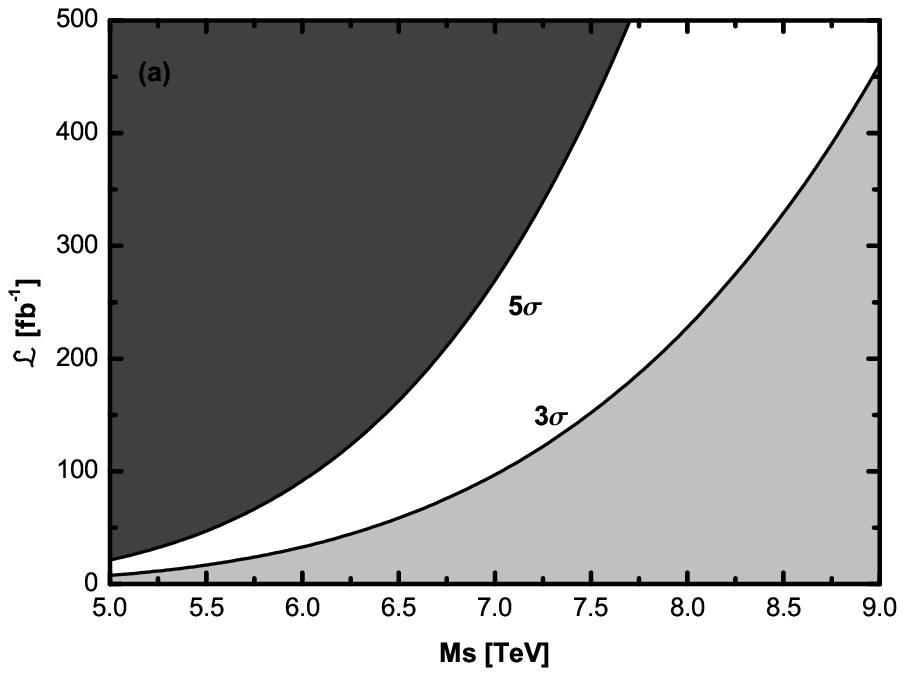}%
\hspace{0in}%
\includegraphics[scale=0.7]{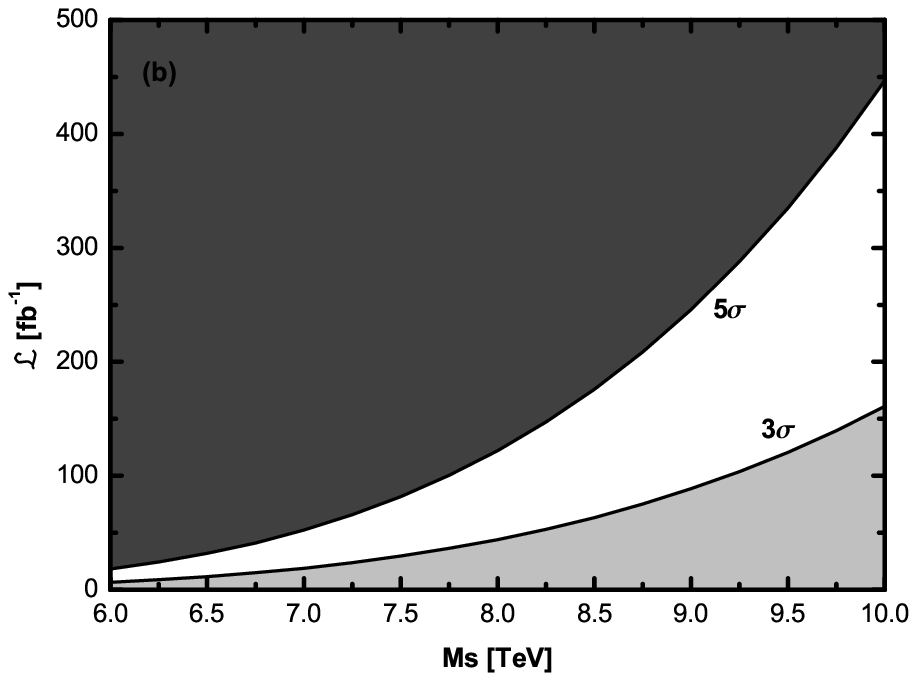}%
\hspace{0in}%
\caption{ \label{fig19} The LED effect discovery area (dark) and
exclusion area (gray) in the $\mathcal{L}-M_S$ space at the
$\sqrt{s}=14$~TeV LHC, where $n=3$. (a) for the $pp \to
W^{+}W^{-}\gamma$ process. (b) for the $pp \to W^{+}W^{-}Z$ process.
}
\end{center}
\end{figure}

From the above analysis, we see that the LED effects can generally
enhance the kinematic observables, especially at the LHC. To
further explore the discovery and exclusion potential for the LED
signals at the ILC and LHC, we adopt the $5\sigma$ discovery
limit and $3\sigma$ exclusion limit to study the constraints on the
fundamental scale $M_S$, namely,
\begin{eqnarray}
\label{upper} \Delta\sigma =\sigma_{LED}-\sigma_{SM} \geq
\frac{5\sqrt{\mathcal{L}\sigma_{LED}}}{\mathcal{L}}\equiv 5\sigma,
\end{eqnarray}
\begin{eqnarray}
\label{lower} \Delta\sigma =\sigma_{LED}-\sigma_{SM} \leq
\frac{3\sqrt{\mathcal{L}\sigma_{LED}}}{\mathcal{L}}\equiv 3\sigma.
\end{eqnarray}
In Figs.\ref{fig18}(a) and \ref{fig18}(b), we present the discovery and
exclusion regions in the $\mathcal{L}-M_{S}$ space for the $e^+e^-
\to W^{+}W^{-}\gamma$ and the $e^+e^- \to W^{+}W^{-}Z$ processes 
at the $\sqrt{s}=800$~GeV ILC with $n=3$, separately. In 
Figs.\ref{fig19}(a) and \ref{fig19}(b), the discovery and exclusion regions for
the $pp \to W^{+}W^{-}\gamma$ and the $pp \to W^{+}W^{-}Z$ processes 
at the $\sqrt{s}=14$~TeV LHC with $n=3$ are given, respectively. 
The dark and gray regions in Figs.18 and 19 represent the $\mathcal{L}-M_{S}$
space with the $5\sigma$ discovery limit and $3\sigma$ exclusion
limit, separately. Some typical limits for $M_S$ with the integrated
luminosity $\mathcal{L}=50$, $100$, $150~fb^{-1}$, which are read
out from Figs.\ref{fig18} and \ref{fig19}, are listed
in Table \ref{tab4} and Table \ref{tab5}, separately. It is found
that the values of the $5\sigma$ discovery and $3\sigma$ exclusion
limits on $M_S$ at the LHC are larger than those obtained at the ILC
with the same integrated luminosity. This reflects the fact that the
$W^{+}W^{-}\gamma/Z$ production rates at the LHC are enhanced by the
additional $gg$ fusion subprocess and the KK-graviton resonant
effect induced by the available large colliding energy, and this feature 
can further be viewed as the advantage of the LHC over the ILC in
exploring the LED signature from the $W^+W^-\gamma/Z$ production
measurements \cite{16}.

\begin{table}[htbp]
\begin{center}
\begin{tabular}{|c|c|c|c|c|}
\hline {}Luminosity$(\mathcal{L})$   &\multicolumn{2}{c|}{$e^+e^-
\to W^+W^-\gamma$}  &\multicolumn{2}{c|}{$e^+e^- \to W^+W^-Z$}\\
\cline{2-5}
{}[fb$^{-1}$]               &$M_{S}[TeV]$($3\sigma$)&$M_{S}[TeV]$($5\sigma$)  &$M_{S}[TeV]$($3\sigma$)&$M_{S}[TeV]$($5\sigma$) \\
\hline  50    &3.25                   &2.87                     &3.42                   &3.02                 \\
\hline  100   &3.54                   &3.12                     &3.72                   &3.29                 \\
\hline  150   &3.73                   &3.28                     &3.91                   &3.45                 \\
\hline
\end{tabular}
\caption{  \label{tab4} The discovery ($\Delta\sigma \geq 5\sigma$)
and exclusion ($\Delta\sigma \leq 3\sigma$) limits on the
fundamental scale $M_S$ for the $e^+e^- \to W^+W^-\gamma$ and
$e^+e^- \to W^+W^-Z$ processes at the $\sqrt{s}=800$~GeV ILC, where
$n=3$. }
\end{center}
\end{table}
\begin{table}[htbp]
\begin{center}
\begin{tabular}{|c|c|c|c|c|}
\hline {}Luminosity$(\mathcal{L})$  &\multicolumn{2}{c|}{$pp \to
W^+W^-\gamma$}  &\multicolumn{2}{c|}{$pp \to W^+W^-Z$}\\
\cline{2-5}
{}[fb$^{-1}$]               &$M_{S}[TeV]$($3\sigma$)&$M_{S}[TeV]$($5\sigma$)  &$M_{S}[TeV]$($3\sigma$)&$M_{S}[TeV]$($5\sigma$) \\
\hline  50    &6.35                   &5.54                     &8.18                   &6.92                 \\
\hline  100   &7.04                   &6.07                     &9.20                   &7.75                 \\
\hline  150   &7.48                   &6.42                     &9.87                   &8.28                 \\
\hline
\end{tabular}
\caption{  \label{tab5} The discovery ($\Delta\sigma \geq 5\sigma$)
and exclusion ($\Delta\sigma \leq 3\sigma$) limits on the
fundamental scale $M_S$ for the $pp \to W^+W^-\gamma$ and $pp \to
W^+W^-Z$ processes at the $\sqrt{s}=14$~TeV LHC, where $n=3$.}
\end{center}
\end{table}

\vskip 5mm
\section{Summary}
\par
In this paper, we study the LED effects induced by the virtual
KK-graviton on the $W^{+}W^{-}\gamma$ and $W^{+}W^{-}Z$ production
processes at both the LHC and ILC. The comparison between the
results for these productions at both colliders are made. We
present the transverse momentum and rapidity distributions of final
particles, and find that the LED contributions remarkably affect the
observables of these processes, particularly in either the high
transverse momentum region or the central rapidity region
($y\sim 0$). We see that the relative LED discrepancy becomes larger
with the increment of the invariant mass of final $W$ pair, and the
integrated cross section in the LED model decreases when $M_S$ goes up
for a fixed $n$ at both the ILC and LHC. The $5\sigma$ discovery
and $3\sigma$ exclusion limits on the fundamental scale $M_S$ are
also obtained at both the ILC and LHC. We also find that the
relative LED discrepancy for the $W^+W^-\gamma/Z$ production at the
LHC is generally larger than that at the ILC due to the additional
LED contributions via $gg$ fusion subprocesses and the KK-graviton
exchanging resonant effect induced by the continues large
colliding energy available in the LHC. From the aspect of effectively
exploring the LED signal in measuring $W^+W^-\gamma/Z$ production,
we conclude that the $W^{+}W^{-}\gamma$ and $W^{+}W^{-}Z$
productions at the LHC could have the distinct advantage over at the
ILC.

\vskip 5mm
\par
\noindent{\large\bf Acknowledgments:} This work was supported in
part by the National Natural Science Foundation of China (Grants
No.11075150, No.11005101), and the Specialized Research Fund for the
Doctoral Program of Higher Education (Grant No.20093402110030).


\vskip 5mm
\begin{thebibliography}{99}

\bibitem{1}
  LHC/LC Study Group Collaboration (G. Weiglein et al.), Phys. Rept. {\bf 426}, 47(2006),
  arXiv:hep-ph/0410364.

\bibitem{2}
  Nima Arkani-Hamed, Savas Dimopoulos, Gia Dvali, Phys. Lett. {\bf B429}, 263(1998),
  arXiv: hep-ph/9803315; Phys. Rev. {\bf D59}, 086004(1999), arXiv:hep-ph/9807344.
  I. Antoniadis, N. Arkani-Hamed, S. Dimopoulos, G. Dvali,
  Phys. Lett. {\bf B436}, 257(1998), arXiv:hep-ph/9804398.

\bibitem{3}
  K. Agashe, N.G. Deshpande, Phys. Lett. {\bf B456}, 60(1999),
  arXiv:hep-ph/9902263.

\bibitem{4}
 N. Agarwal, V. Ravindran, V.K. Tiwari, A. Tripathi, Nucl. Phys. {\bf B830}, 248(2010),
 arXiv:0909.2651; Phys. Rev. {\bf D82}, 036001(2010), arXiv:1003.5450.
 M.C. Kumar, P. Mathews, V. Ravindran, A. Tripathi, Phys. Lett. {\bf B672} 45(2009),
 arXiv:0811.1670.

\bibitem{5}
 S. Ask, Eur. Phys. J. {\bf C60}, 509(2009),
 arXiv:0809.4750. Xiangdong Gao, et al., Phys. Rev. {\bf D81}, 036008(2010),
 arXiv:0912.0199. M.C. Kumar, P. Mathews, V. Ravindran, S. Seth, Nucl. Phys. {\bf B847}, 54(2011),
 arXiv:1011.6199.

\bibitem{5-1}
 Y.-M. Bai, L. Guo, X.-Z. Li, W.-G. Ma, R.-Y. Zhang, Phys. Rev. {\bf D85},016008(2012),
 arXiv:1112.4894.

\bibitem{6}
 S. Karg, M. Kramer, Qiang Li, D. Zeppenfeld, Phys. Rev. {\bf D81}, 094036(2010),
 arXiv:0911.5095.

\bibitem{9}
 S. Godfrey, 'Quartic Gauge Boson Couplings', arXiv:hep-ph/9505252. O.J.P. Eboli,
 M.C. Gonzalez-Garcia, S.M. Lietti, Phys. Rev. {\bf D69}, 095005(2004),  arXiv:hep-ph/0310141.

\bibitem{10}
 F. Ferro, et al., To appear in the proceedings of Forward Physics at Conference: C10-05-27, arXiv: 1012.5169

\bibitem{7}
 A. Lazopoulos, K. Melnikov, F. Petriello, Phys. Rev. {\bf D76}, 014001(2007),
 arXiv: hep-ph/0703273. V. Hankele, D. Zeppenfeld, Phys. Lett. {\bf B661}, 103(2008),
 arXiv:0712.3544. G. Bozzi, et al., Phys. Rev. {\bf D81}, 094030(2010),
 arXiv:0911.0438.

\bibitem{8}
 M.C. Kumar, P. Mathews, V. Ravindran, S. Seth, Phys. Rev. {\bf D85}, 094507(2012),
 arXiv:1111.7063.

\bibitem{11}
 ALEPH and DELPHI and L3 and OPAL and LEP Electroweak Working Group Collaborations, CERN-PH-EP/2005-051,
 arXiv: hep-ex/0511027. Measurement of the W+W-gamma Cross Section and Direct Limits on Anomalous Quartic
 Gauge Boson Couplings at LEP L3 Collaboration, Phys. Lett. {\bf B490}, 187(2000),
 arXiv:hep-ex/0008022.

\bibitem{13}
 Oscar J.P. Eboli, M.C. Gonzalez-Garcia, S.M. Lietti, S.F. Novaes, Phys. Rev. {\bf D63}, 075008(2001),
 arXiv:hep-ph/0009262. Dan Green(Fermilab), arXiv:hep-ex/0310004.

\bibitem{14}
 Jurgen Reuter, ECONF C0705302: TEV04, 2007, FR-THEP-07-02, arXiv:0708.4383v1.
 Hitoshi Yamamoto, J. Phys. Soc. Jap. {\bf 76}, 111014(2007), arXiv:0709.0899v1.

\bibitem{15}
 Tao Han, Joseph D. Lykken, Ren-Jie Zhang, Phys. Rev. {\bf D59}, 105006(1999),
 arXiv:hep-ph/9811350.

\bibitem{16}
 Gian F. Giudice, R. Rattazzi, James D. Wells, Nucl. Phys. {\bf B544}, 3(1999),
 arXiv:hep-ph/9811291.

\bibitem{19}
 J. Pumplin, D.R. Stump, J. Huston, H.L. Lai, Pavel M. Nadolsky, W.K. Tung,  JHEP {\bf 0207}, 012(2002),
 arXiv:hep-ph/0201195.

\bibitem{20}
 J. Beringer et al. (Particle Data Group), Phys. Rev. {\bf D86}, 010001(2012).

\bibitem{21}
 ATLAS Collaboration (Georges Aad (Freiburg U.) et al.), CERN-PH-EP-2011-189, Phys. Lett. {\bf B710}, 538(2012),
 arXiv: 1112.2194.

\bibitem{22}
 CMS Collaboration, CMS-EXO-11-038; CERN-PH-EP-2011-173,
 arXiv:1112.0688v1.

\bibitem{23}
 CMS Collaboration, arXiv:1202.3827v2.

\bibitem{17}
 T. Hahn, Comput. Phys. Commun. {\bf 140}, 418(2001),
 arXiv:hep-ph/0012260.

\bibitem{24}
 E.Boos, et al., PoS ACAT08:008,2009, arXiv:0901.4757.

\end{thebibliography}
\end{document}